\documentclass{PoS}
\usepackage{amsmath}
\usepackage{amstext}
\usepackage{amsthm}
\usepackage{amssymb}
\usepackage{graphicx}
\usepackage{mathrsfs}
\usepackage[bottom]{footmisc}
\usepackage{indentfirst}
\usepackage{upref}
\usepackage{cite}
\usepackage{units}
\usepackage{mciteplus}
\def\rp{r_{+}}
\def\re{r_{\rm EH}}
\newcommand{\be}{\begin{equation}}
\newcommand{\ee}{\end{equation}}
\newcommand{\ba}{\begin{eqnarray}}
\newcommand{\ea}{\end{eqnarray}}
\newcommand{\bc}{\begin{center}}
\newcommand{\ec}{\end{center}}

\newcommand{\bay}{\begin{array}{rcl}}
\newcommand{\eay}{\end{array}}

\def\ie{{i.e. }}
\def\lp{\ell_{\rm Pl}}
\def\LH{\ell_{\rm H}}

\def\tr{t_{\rm tr}}
\def\mp{m_{\rm Pl}}
\def\tp{t_{\rm Pl}}
\def\OM{\Omega_{\rm M}}
\def\OL{\Omega_{\Lambda}}

\def\OLS{\Omega_\Lambda^\ast}

\def\barg{\bar{G}}
\def\lat{\lambda_{\rm T}}

\def\kat{k_{\rm T}}
\def\gat{g_{\rm T}}

\def\h0{H_{0}}

\def\calp{\widetilde{\cal P}} 
\def\calp{\widetilde{\mathscr P}} 
\def\clp{{\mathscr P}} 
 
\def\UU{{\cal U}} 
\def\ssc{S_{\rm c}}

\title{Astrophysical implications of the Asymptotic Safety Scenario in Quantum
Gravity}

\ShortTitle{Asymptotic Safety in Astrophysics}

\author{\speaker{Alfio Bonanno}\\
INAF, Osservatorio Astrofisico di Catania, Via S.Sofia 78, 95123 Catania,
Italy.\\ INFN, Via S.Sofia 64, 95123 Catania, Italy.\\
        E-mail: \email{abo@oact.inaf.it}}
\abstract{
In recent years it has emerged that the high energy behavior
of gravity could be governed by an ultraviolet non-Gaussian fixed point of the
(dimensionless) Newton's constant, whose behavior at high energy is thus {\it
antiscreened}. This phenomenon has several astrophysical implications.  
In particular in this article recent works on renormalization group  improved
cosmologies based upon a renormalization group  trajectory of Quantum Einstein Gravity with
realistic parameter values will be reviewed. It will be argued
that quantum  effects can account for the entire entropy of the present
Universe in the massless sector and give rise to a phase of inflationary expansion.
Moreover the prediction for the final
state of the black hole evaporation is a Planck size remnant which is formed
in an infinite time. }

\FullConference{Workshop on Continuum and Lattice Approaches to Quantum Gravity\\
                 September 17-19 2008\\
                 Brighton, University of Sussex, United Kingdom}

\begin{document}

\section{Introduction}
Cosmology is a natural setting to study quantum gravity,
which  may provide answers to fundamental
questions as why is the expansion of the universe isotropic, can the initial singularity be avoided, 
why does the vacuum energy ``gravitate'' so little (Cosmological Constant
problem)?

%The gist of the key problems in modern cosmology is clearly
%linked to the physical meaning of Newton's constant and the Cosmological
%constant, whose smallness can be rephrased as why the vacuum energy gravitates
%so little.

%One example is provided by the so called ``flatness problem''  which can 
%be rephrased  in term of the ``Machian'' equality
%\be\label{mach}
%(G\rho t^2)_{\rm today} = O(1)
%\ee
%where $\rho$ is  the observed matter density of the Universe,  $t$ the age of
%the Universe and $G$ the Newton constant. One is thus tempted to speculate that
%a more natural framework to explain Eq.(\ref{mach}) is to argue that $G$ is a
%dynamically varying quantity and not a ``constant''.

In recent years it has emerged that the asymptotic safety
scenario \cite{wein2,2009AnPhy.324..414C,niedermaier_asymptotic_2006}
could provide the right framework to address the above questions. 
According to this approach the  ultraviolet (UV) behavior of
quantum gravity is controlled by a fixed point 
at a non-zero value of the (dimensionless) coupling constant, so that the
dimensionful Newton's constant reduces its strength at higher energies, 
it is thus {\it antiscreened}. The non-perturbative
renormalization group (RG) equation employed in this investigation predicts
that the dimensionless cosmological constant reaches a non-gaussian fixed
 point (NGFP) in the infinite cutoff limit, so that the full Einstein-Hilbert
 Lagrangian is renormalizable at a non-perturbative level around this fixed
 point. 

The gravitational {antiscreening} behavior is very 
similar to the running of the non-Abelian gauge coupling in Yang-Mills
Theory, but only after the introduction of the effective average action 
and its functional renormalization group equation for gravity
\cite{1998PhRvD..57..971R} detailed investigations of the scaling
behavior of the Newtons's constant have become possible 
\cite{1998PhRvD..57..971R,1998CQGra..15.3449D,2002PhRvD..65b5013L,2002PhRvD..66b5026L,2002CQGra..19..483L,
2002PhRvD..65f5016R,2002PhRvD..66l5001R,1999PThPh.102..181S,2003PhRvD..67h1503P,2004PhRvL..92t1301L,
2005JHEP...02..035B,2006PhRvL..97v1301C,2009GReGr..41..983R,2009PhRvD..79j5005R,2009arXiv0904.2510M,2009arXiv0905.4220M}.
The non-perturbative renormalization group equation underlying this approach defines a Wilsonian
RG flow on a theory space which consists of all diffeomorphism invariant
functionals of the metric $g_{\mu\nu}$. 

This framework  turned out to be an ideal setting for investigating the
asymptotic safety scenario in gravity \cite{wein2,2009AnPhy.324..414C,niedermaier_asymptotic_2006} and, in fact,
substantial evidence was found for the non-perturbative renormalizability of Quantum Einstein Gravity.
The theory emerging  from this construction (``QEG") is not a quantization of
classical general relativity. Instead, its bare action corresponds to a nontrivial 
fixed point of the RG flow and therefore is a {\it prediction}.  The
effective average action \cite{1998PhRvD..57..971R,2002PhR...363..223B} has crucial advantages 
as compared to other continuum implementations of the Wilson RG, in particular it is closely related to the standard 
effective action and defines a family of effective field theories
$\{ \Gamma_k[g_{\mu\nu}], 0 \leq k < \infty \}$ labeled by the coarse graining scale $k$.
The latter property opens the door to a rather direct extraction of physical
information from the RG flow, at least in single-scale cases: If the physical process 
or phenomenon under consideration involves only a single typical momentum
scale $p_0$ it can be described by a tree-level evaluation of $\Gamma_k[g_{\mu\nu}]$, with $k=p_0$.
The precision which can be achieved by this effective field theory description
depends on the size of the fluctuations relative to the mean values. If they are large, or if more than
one scale is involved, it might be necessary to go  beyond the tree analysis. 

The qualitative scale dependence of Newton's constant can be grasped with the
help of the following physical argument. 
Let us imagine that in the large distance limit the leading quantum effects of
the geometry are described by quantizing the linear fluctuations of the metric, $g_{\mu \nu}$. 
The resulting theory is a minimallu coupled theory in a
curved background spacetime whose elementary quanta, the gravitons, carry energy and momentum. 
The vacuum of this theory will be populated by virtual graviton pairs, and the
problem is to understand how these virtual gravitons respond to the perturbation by an external
test body which we immerse in the vacuum. Assuming that also in this
situation gravity is universally attractive, the gravitons will be
attracted towards the test body. It will thus become ``dressed" by a
cloud of virtual gravitons surrounding it so that its effective mass
seen by a distant observer is larger than it would be in absence of
any quantum effects. This means that while in QED the quantum
fluctuations {\it screen} external charges, in quantum gravity they
have an {\it antiscreening} effect on external test masses. The consequence of this simple 
{\it Gedanken} experiment entails Newton's constant becoming a scale dependent
quantity $ G\!\left(k \right)$ which is small at small distances $ r\sim 1/k$, 
and which becomes large at larger distances. 

In QED the {screening} behavior is well-known but it is interesting to recall 
how this result is obtained from the "renormalization group improvement",   
a standard device, in particle physics, in order to add the dominant quantum corrections 
to the Born approximation of a scattering cross section for instance. 
One starts from the classical potential energy $V_{\rm cl}(r) = e^2/4\pi r$ and replaces $e^2$ by the running
gauge coupling in the one-loop approximation: 
\begin{equation}\label{uel}
e^2(k)=e^2(k_0)[1-b\ln(k/k_0)]^{-1},\;\;\;\;\; b\equiv e^2(k_0)/6\pi^2.
\end{equation}
The crucial step is to identify the renormalization point $k$ with the inverse of the 
distance $r$ 
%This is possible because in the massless theory $r$
%is the only dimensionful quantity which could define a scale.
so that result of this substitution reads
\be\label{rea}
V(r)=-e^2(r_0^{-1})[1+b\ln(r_0/r) + O(e^4)]/4\pi r
\ee
where the IR reference scale $r_0\equiv 1/k_0$ has to be kept finite in the
massless theory. We emphasize that eq.(\ref{rea}) is the correct (one-loop, massless)
Uehling potential which is usually derived by more conventional perturbative methods
\cite{dittrich_effective_1985}. Obviously the position dependent
renormalization group improvement $e^2\rightarrow e^2(k)$, $k\propto 1/r$ encapsulates
the most important effects
which the quantum fluctuations have on the electric field produced by a point
charge. 

%The  Maxwell equations now become
The effective field theory techniques proved useful for an understanding 
of the scale dependent geometry of the effective QEG spacetimes
\cite{2005JHEP...10..050L,2006JHEP...01..070R,2007JHEP...01..049R}. In
particular it has been shown \cite{2002PhRvD..65b5013L,2005JHEP...10..050L} that these spacetimes
have fractal properties, with  a fractal dimension of 2 at small, and 4 at large distances. The same dynamical dimensional reduction was also observed
in numerical studies of Lorentzian dynamical triangulations 
\cite{2005PhLB..607..205A,2004PhRvL..93m1301A,2005PhRvL..95q1301A} and in
\cite{2006JHEP...11..081C} A.Connes et al. speculated about 
its possible relevance to the non-commutative geometry of the standard model.

In order to extract all the relevant information from the RG evolution, it is
thus necessary to relate the cutoff scale $k$ which corresponds to the
resolution of the RG flow,  to the spacetime properties.  
This  procedure is called ``cutoff identification" for which the relevant energy scale $k$ is related 
to a characteristic length scale where
the quanta with energy $k$ propagate.  In the case of massless QED the 
choice $k\propto 1/r$ was clearly the only possible one, as there are no other
relevant scales in the problem. When several scales are present the
prescription which emerges from the general theory of the Effective Average
Action \cite{2002PhR...363..223B} is that $\Gamma_k$  
is defined at a scale $k$ which is the {\it largest} one
of the various competing scales in the fluctuation 
determinant of the Average Action, $\Gamma^{(2)}_k$, 
namely
\be\label{flucd}
\Gamma^{(2)}_k=\frac{\delta^2 \Gamma_k}{\delta \Phi^2 }
\ee
where $\Phi$ is the so-called ``blocked''  field \cite{1995PhRvD..52..969B}.  

The  difficulty arises when we decide to apply the same
``recipe'' in gravity by writing
\be\label{klen}
k \sim 1/\ell(x^\mu)), \;\;\;\;\;\;\;\;\;\;\;   \ell=\ell(g_{\mu\nu})  
\ee
being $\ell$ a characteristic length where the fluctuations with energy $k$
propagate.  The reason is that the flow equation is by construction
diffeomorphism invariant at any $k$ so that the RG flow itself does 
not know anything about the background field metric $g_{\mu\nu}$ that has been
used for projecting on a finite-dimensional subspace of the ``theory space''.

There are two possible strategies to overcome this issue. The first one amounts
to choose a fiducial metric which is a {\it solution} of the Einstein equations
and RG-improve it by substituting the Newton constant $G$ with the running
$G(k)$ together with a cutoff identification of the type (\ref{klen}). 
The limitation of this approach lies in the fact that in general the improved
metric may not be a solution of Einstein equation, but 
one can imagine that this is a sort of ``Thomas-Fermi'' 
approximation where only the leading quantum corrections are  taken into account
\cite{2000PhRvD..62d3008B,2006PhRvD..73h3005B}. The improved $g_{\mu\nu}(k)$
metric represents then a sort of ``emergent'' spacetime description of the
effective geometry
\cite{2005LRR.....8...12B,2004CQGra..21.1725M,2007arXiv0712.0810V}  according 
to the scale dependence of the Newton constant. 

A second possibility is to consider the energy scale $k$ associated to the
field strength itself rather than to an observational scale $\ell$. This is
motivated by the analogy with the QED (and QCD) case, where higher loop 
contributions to the Uehling potential are obtained by renormalization group 
improvement of the QED {\it action} by using the field strength
$(F_{\mu\nu}F^{\mu\nu})^{1/4}$  as a cutoff instead than $1/r$
\cite{1973NuPhB..52..483M,1978NuPhB.134..539M,1978NuPhB.143..485P}.
In this case the short distance correction to the static potential is obtained
from the non-linear differential equations
\begin{eqnarray}\label{maty}
&&\nabla \cdot  \mathbf D = J_0, \;\;\;\;\; \mathbf D = \mathbf E \; \epsilon
(E) , \;\;\; \mathbf E = -\nabla A^0\\[2mm]\nonumber
&& \epsilon(E)=1-\frac{e^2}{12\pi^2} \log (e E/k_0^2)+\ldots
\end{eqnarray}
whose solution reproduces the Uehling potential in the long distance limit, but
in general the solutions of Eq.(\ref{maty})  include higher loop effects
due to the non-linearities of the effective action in the short distance limit.

The two approaches discussed above are obviously related, at least in some
limit. In the case of Robertson-Walker spaces it will be shown that due to the
very high degree of symmetry of the spacetime, the time-scale defined by ``Hubble
parameter''  behaves essentially like the characteristic time scale
associated to the relevant curvature
invariants\cite{2004ForPh..52..650R,2005JCAP...09..012R,2007JCAP...08..024B}. 
In the case of spherically symmetric spacetimes
\cite{2000PhRvD..62d3008B,2004JCAP...12..001R,2004PhRvD..70l4028R,2006PhRvD..73h3005B},
near the singularity the proper distance of a radially free falling
observer behaves essentially as $1/\sqrt{\Psi_2}$, being $\Psi_2$ the
``Coulombian'' component of the Weyl tensor. 

From the above discussion it is then clear that in general there is not
a preferred strategy to perform the RG improvement in gravity. In some case it
might  be  more interesting to RG improve {\it  solutions} and to
make contact with an emergent spacetime description of the effective geometry. In
some other cases it could be more convenient to work with a RG improvement at the level of  
{\it field equations} or {\it actions}
\cite{2004PhRvD..69j4022R,2004PhRvD..70l4028R,2004CQGra..21.5005B,2005IJMPA..20.2358B}.

It is important to remark that it is not surprising  that
different cutoff might provide quantitavely different evolutions, as usually the
$\beta$-functions are not ``universal'' quantities. For instance it is well
known \cite{Goldenfeld:1992qy} that different realizations of the block-spin RG 
transformation applied to the Ising
model may provide different values for fixed points, as we are essentially
using different type of ``microscopes''\footnote{From this point of view the
criticism expressed in \cite{2008PhRvL.101h1301W} should not be seriously considered.}.
On the other hand truly universal quantities, like the critical 
exponents, are essentially insensitive to the cutoff choice.

In this review recent results obtained with the RG improvement of Einstein
theory will be discussed in the framework QEG will be reviewed, with particular
emphasis on recent results obtained in cosmology \cite{2007JCAP...08..024B}.
In particular in Sec.2 the RG evolution of the Newton constant and Cosmological
constant describing our Universe are described. In Sec.3 a covariant formalism
to improve the Einstein field equation is presented while in Sec.4 the RG
improved Robertson-Walker Cosmology is discussed. In Sec.5 the basic
mechanism to produce the entropy of the Universe is presented. In Sec.6 the
properties of a class of solutions of the RG equations are discussed. 
In Sec.7 a mechanism to produce a power-law inflation is studied.
In Sec.8 and Sec.9 the properties of RG improved Black Hole metric is studied. 
In Sec.10 the possibility that Quantum Gravity effects are present on
Astrophysical distances is reviewed. Sec.11 is devoted to the Conclusions.

\section{The RG trajectory of our Universe}
It is possible to show that there exists a class of RG trajectories
obtained from QEG in the Einstein-Hilbert approximation \cite{1998PhRvD..57..971R}, 
namely those of the ``Type IIIa''
\cite{2002PhRvD..65f5016R} which possesses all the qualitative properties one
would expect from the  RG trajectory describing gravitational phenomena
in the real Universe we live in. In particular they can have a long classical regime and a small,
positive cosmological constant in the infrared. Determining its parameters from observations,
one finds \cite{2007JCAP...08..024B} that,  according to this particular QEG trajectory, the
running cosmological constant $\Lambda(k)$ changes by about 120 orders of magnitude between $k$-values of the order
of the Planck mass and macroscopic scales, while the running Newton constant $G(k)$ has no
strong $k$-dependence in this regime. For $k> \mp$, the non-Gaussian fixed point 
which is responsible for the renormalizability of QEG controls their scale dependence. In the deep
ultraviolet $(k\rightarrow \infty)$, $\Lambda(k)$ diverges and $G(k)$ approaches zero.

Is there any experimental or observational
evidence that would  hint at this enormous scale dependence of the gravitational parameters,
the cosmological constant in particular?
As it was stressed before, even though  it is always difficult to give a
precise physical interpretation to the RG scale $k$ it is fairly 
certain that any sensible identification of $k$ in terms
of cosmological quantities will lead to a $k$ which decreases during the expansion
of the Universe. As a consequence, $\Lambda(k)$ will also decrease as the Universe expands.
Already the purely qualitative assumption of a {\it positive} and {\it decreasing}
cosmological constant supplies an interesting hint as to which
phenomena might reflect a possible $\Lambda$-running.
\begin{figure}[h]
\begin{center}$
\begin{array}{cc}
\includegraphics[width=2.in]{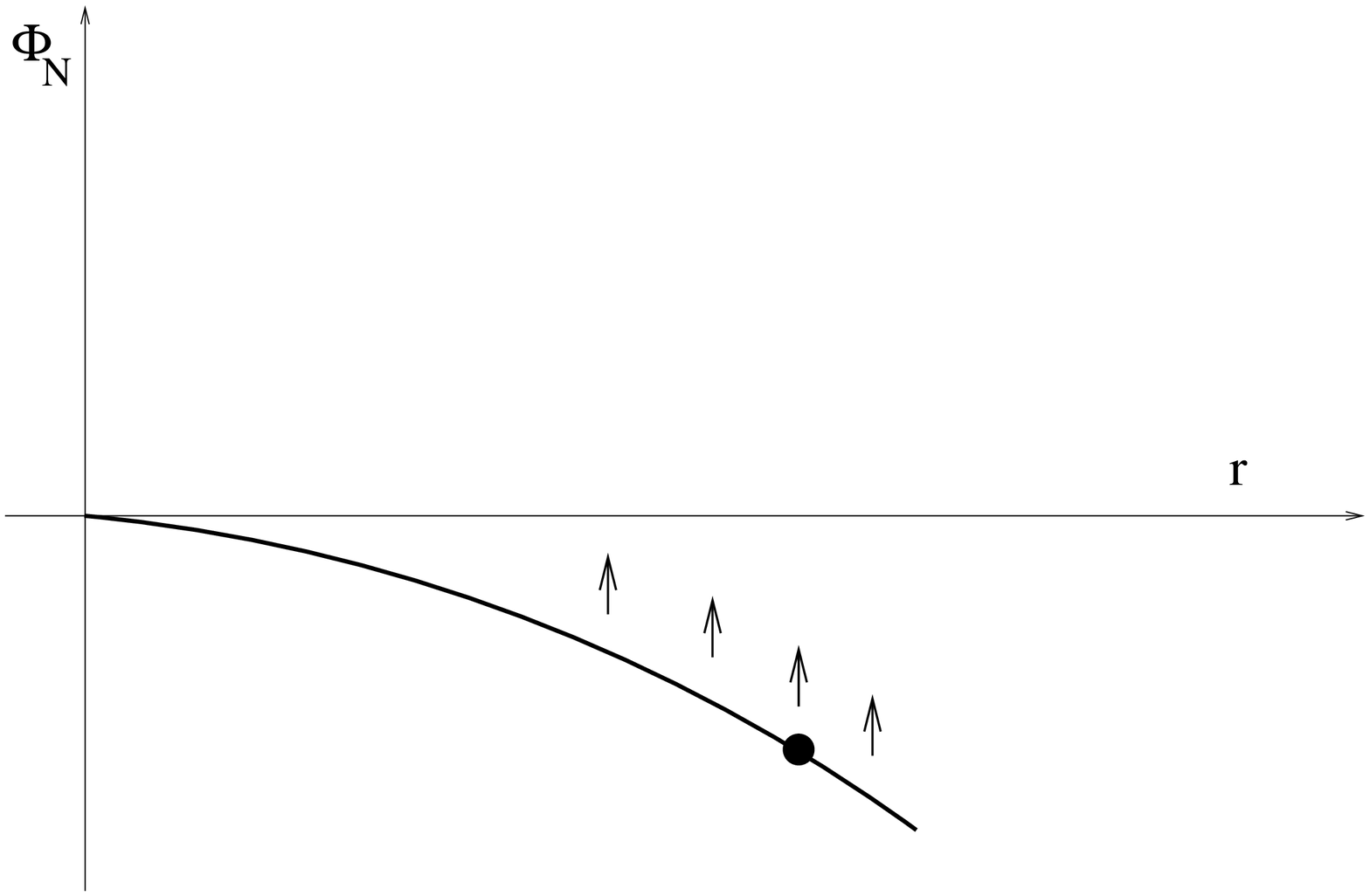}&
\includegraphics[width=3.in]{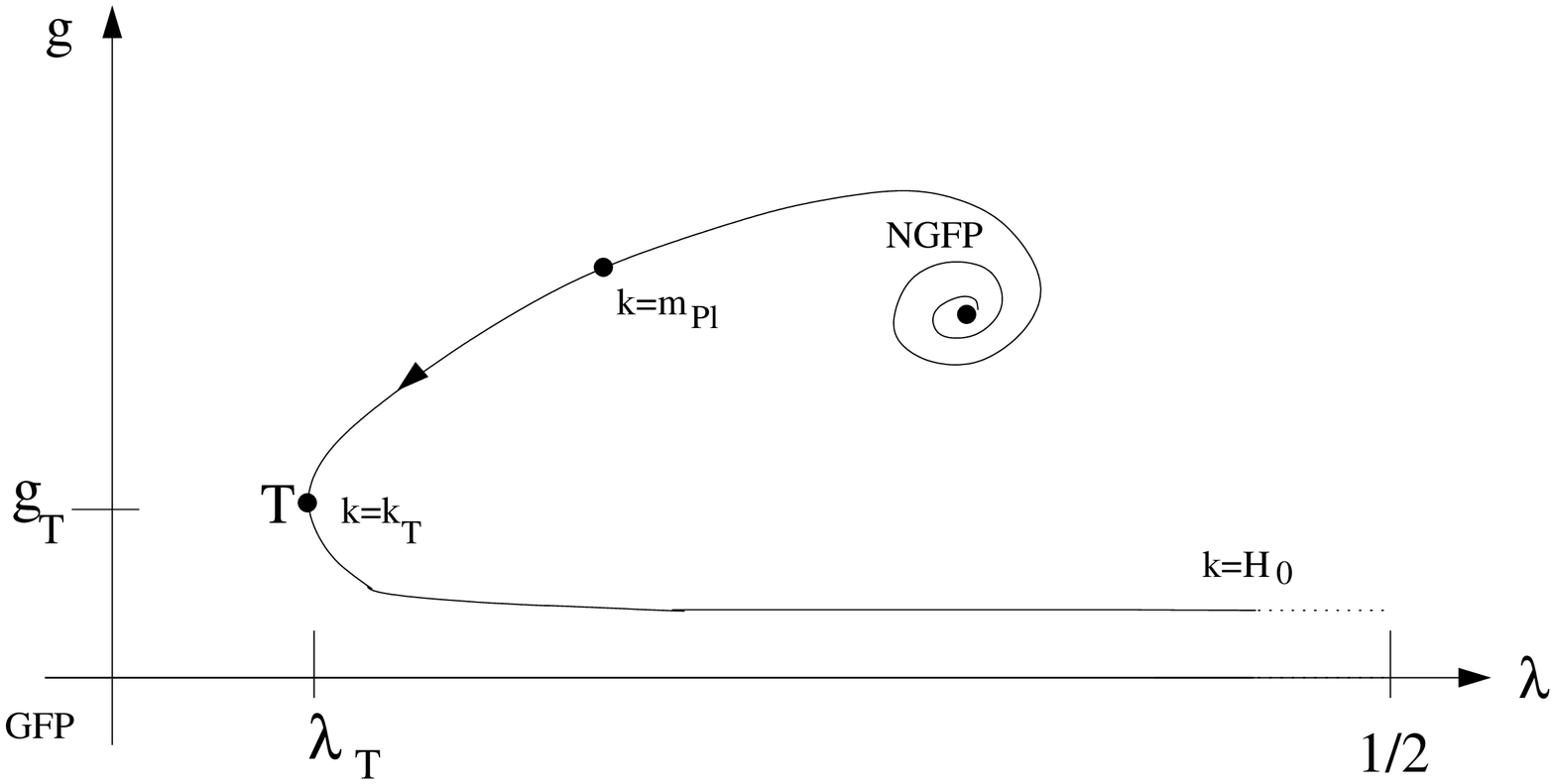}
\end{array}$
\end{center}
\caption{ The left panel shows the quasi-Newtonian potential corresponding to de Sitter space.
The curve moves upward as the cosmological constant decreases.
On the right panel the ``realistic" RG trajectory.}
\label{fig1}
\end{figure}
                         
To make the argument as simple as possible, let us first consider a Universe without matter,
but with a positive $\Lambda$. Assuming maximal symmetry, this is nothing but  de Sitter
space, of course. In static coordinates its metric is
%\be\label{1.1}
$ds^2 = -(1+2\Phi_{\rm N}(r) ) dt^2+ (1+2\Phi_{\rm N}(r))^{-1}dr^2 +
r^2 (d\theta^2 +\sin^2 \theta d\phi^2)$
%\ee
with
%\be\label{1.2}
$\Phi_{\rm N}(r) = -\frac{1}{6}\; \Lambda\; r^2$.
%\ee
In the weak field and slow motion limit $\Phi_{\rm N}$ has the interpretation
of a Newtonian potential, with a correspondingly simple physical interpretation.
The left panel of Fig.1 shows $\Phi_{\rm N}$ as a function of $r$; for $\Lambda >0$ it is
an upside-down parabola. Point particles in this spacetime, symbolized by the
black dot in Fig.1, ``roll down the hill'' and are rapidly driven away
from the origin and from any other particle. Now assume that the magnitude of $|\Lambda|$
is slowly (``adiabatically'') decreased. This will cause the potential $\Phi_{\rm N}(r)$
to move upward as a whole, its slope decreases. So the change in $\Lambda$ increases the particle's
potential energy.  This is the simplest way of understanding that a {\it positive decreasing}
cosmological constant has the effect of ``pumping'' energy into the matter degrees of freedom.
More realistically one will describe the matter system in a hydrodynamics or quantum field
theory language and one will include its backreaction onto the metric. But the basic conclusion,
namely that a slow decrease of a positive $\Lambda$ transfers energy into the matter system, will
remain true.

We are thus led to  suspect that, because of the decreasing cosmological constant,
there is a continuous inflow of energy into the cosmological fluid contained in an
expanding Universe. It will ``heat up'' the fluid or, more exactly, lead to a slower decrease
of the temperature than in standard cosmology. Furthermore, by elementary thermodynamics, it will
{\it increase} the entropy of the fluid. If during the time $dt$ an amount of heat
$d Q>0$ is transferred into a volume $V$ at the temperature $T$ the entropy changes by an amount $dS=dQ/T>0$.
To be as conservative (i.e., close to standard cosmology) as possible, we assume that this process
is reversible. If not, $dS$ is even larger.

In standard Friedmann-Robertson-Walker (FRW) cosmology the expansion is adiabatic,
the entropy (within a comoving volume) is constant. It has always been somewhat puzzling therefore
where the huge amount of entropy contained in the present Universe comes from.
Presumably it is dominated by the CMBR photons which  contribute an amount
of about $10^{88}$ to the entropy within the present Hubble sphere. (We use units such that
$k_{\rm B}=1$. ) In fact, if it  true that no entropy is produced during the
expansion then the Universe would have had an entropy of at least $10^{88}$ immediately
after the initial singularity which for various reasons seems quite unnatural.
In scenarios which invoke a ``tunneling from nothing'', for instance,
spacetime was ``born'' in a pure quantum state, so the very early Universe is expected to have
essentially no entropy. Usually it is argued that the present entropy 
is the result of some sort of ``coarse graining'' which, however, typically 
is not considered as an active part of the
cosmological dynamics in the sense that it would have an impact on the
time evolution of the metric.

In \cite{2007JCAP...08..024B}   it was argued that in principle the entire
entropy of the massless fields in the present universe can be understood 
as arising from the mechanism described above.
If energy can be exchanged freely between
the cosmological constant and the matter degrees of freedom, the entropy observed today
is obtained precisely if the initial entropy at the ``big bang'' vanishes.
The assumption that the matter system must allow for an unhindered energy exchange with
$\Lambda$ is essential. 

There is another, more direct potential consequence of a decreasing positive cosmological constant,
namely a period of automatic
inflation during the very first stages of the cosmological evolution. It is not surprising,
of course, that a positive $\Lambda$ can cause an accelerated expansion, but in the
classical context the problem with a $\Lambda$-driven inflation is that it would never
terminate once it has started. In popular models of scalar driven inflation
this problem is circumvented by designing the inflaton potential in such a way that it
gives rise to a vanishing vacuum energy after a period of ``slow roll''.

There will thus be reviewed generic RG cosmologies based upon the QEG
trajectories which have an era of $\Lambda$-driven inflation immediately after
the big bang which ends automatically as a consequence of the RG running of $\Lambda(k)$. 
Once the scale $k$ drops significantly below
$\mp$, the accelerated expansion ends because the vacuum energy density $\rho_\Lambda$ is already
too small to compete with the matter density. Clearly this is a very attractive scenario:
{\it no ad hoc ingredients such as an inflaton field or a special potential
are needed to trigger inflation}. It suffices to include the leading quantum effects in the gravity + matter
system. Furthermore, it will be shown that asymptotic safety offers a natural mechanism
for the quantum mechanical generation of primordial density perturbations, 
the seeds of cosmological structure formations.

\section{A covariant approach to RG improvement in cosmology}
In the following we shall present the improved RG equation in the $3+1$
formalism. Let $g_{\mu\nu}$ be the space-time metric with signature $(-,+,+,+)$.
A ``cosmological fundamental observer'' comoving with the cosmological fluid
has 4-velocity $u^\mu= dx^\mu/ d\tau$ with $ u^\mu u_\mu = -1$,
where $\tau$ is the proper time along the fluid flow lines.
The projection tensor onto the tangent 3-space orthogonal to $u^\mu$ is
$h_{\mu\nu} = g_{\mu\nu}+u_\mu u_\nu$,
with  ${h^\mu}_\nu {h^{\nu}}_\sigma =
{h^\mu}_\sigma$ and ${h^{\mu}}_\nu u^\nu =0$.
We denote by a semicolon the standard covariant derivative and by an over-dot
the differentiation with respect to the proper time $\tau$.
The covariant derivative of $u^{\mu}$ reads as
\be\label{2.3}
u_{\mu;\nu}=\omega_{\mu\nu}+\sigma_{\mu\nu}+
\frac{1}{3}\Theta h_{\mu\nu}-\dot{u}_{\mu} u_{\nu},
\ee
where $\omega_{\mu\nu}={h^{\alpha}}_\mu {h^{\beta}}_\nu u_{[\alpha;\beta]}$
is the vorticity tensor, $\sigma_{\mu\nu} = {h^{\alpha}}_\mu {h^{\beta}}_\nu
u_{(\alpha;\beta)} -\frac{1}{3}\Theta h_{\mu\nu}$
is the shear tensor, $\Theta = {u^\mu}_{;\mu}$ is the expansion scalar and
$\dot{u}^\mu = {u^{\mu}}_{;\nu} u^\nu$ is the acceleration
four-vector; square and round brackets denote anti-symmetrization and
symmetrization, respectively. 

One can introduce a representative length $\ell$ along the particle world-lines
by the equation
\be\label{lun}
{\dot{\ell}}/{\ell}=\frac{1}{3}\Theta
\ee
In fact $\ell$ represent completely the volume behavior of the fluid as any
comoving volume element is proportional to $\ell^3$. The net effect of $\Theta$
is in fact to change fluid sphere into another fluid sphere with the the same
orientation but with different volume. One can then define the ``Hubble
parameter'' and the deceleration parameter by
\be\label{hubble}
H\equiv \dot \ell /\ell, \;\;\;\;\;\; q\equiv \ddot \ell / \ell H^2
\ee
The Einstein equations  read
\be\label{2.4}
%R_{\mu\nu}-{1\over 2} R g_{\mu\nu}= -\Lambda g_{\mu\nu}+8\pi G \; T_{\mu\nu},
R_{\mu\nu}u^\mu u^\nu =  4 \pi G(\rho +3 p)-\Lambda, \;\;\;\;
R_{\alpha\beta}{h^{\alpha}}_\mu{h^{\beta}}_\nu = 
\Big [ 4\pi G(\rho - p) +\Lambda \Big] 
\ee
where $\Lambda=\Lambda(x^\mu)$ is the position-dependent cosmological term and
$G=G(x^\mu)$ the position-dependent Newton parameter.
The energy-momentum tensor is assumed to have the 
the  perfect fluid form
$T^{\mu\nu} = (p+\rho) \; u^\mu u^\nu + p\;g^{\mu\nu}$.
The Bianchi identities lead to the conservation law along $u^\mu$
\be\label{2.6}
{\dot{\rho}} +\Theta(\rho +p)=-\frac{1}{8\pi G} \Big [8\pi \dot G \rho
+\dot\Lambda \Big ] ,
\ee
and onto the orthogonal hypersurface
\be\label{2.7}
\dot{u}^\mu + {h^{\mu\nu}p_{;\nu}\over \rho + p} =
\frac{h^{\mu\nu}}{8\pi G} \Big [\Lambda_{,\nu} - 8 \pi p G_{,\nu} \Big]
\ee

The {\it Raychaudhuri equation} is obtained with the help of the Einstein field
equations and of Eq. (\ref{2.3}), 
\be\label{2.9}
\dot{\Theta}+{1\over 3}\Theta^2+2(\sigma^2-\omega^2)-
\dot{u}^\mu_{\;\; ;\mu}+4 \pi G (\rho+3p)-\Lambda=0,
\ee
where $2\sigma^2 \equiv  \sigma_{\mu\nu}\sigma^{\mu\nu}$ and
$2\omega^2 \equiv \omega_{\mu\nu}\omega^{\mu\nu}$.
%The term $ \dot{u}^\mu_{\;\; ;\mu}$
%is identically vanishing for homogeneous spaces (\ldots).
The scalar curvature of the tangent space is given by
\be\label{2.10.0}
{\cal K}\equiv {}^{(3)}R = R+2R_{\mu\nu}u^{\mu}u^{\nu}
+2\sigma^2-2\omega^2-{2\over 3}\Theta^{2},
\ee
which leads, by using the field equations (\ref{2.4}),
to the {\it generalized Friedmann equation}
\be\label{2.10}
{\cal K}= 2\sigma^2 -2\omega^2 -{2\over 3}\Theta^2
+16\pi G\rho +2\Lambda.
\ee
In homogeneous spaces, Eq.(\ref{2.7})  is identically satisfied, while
Eq.(\ref{2.10}) reduces to the familiar  Friedmann equation which is coupled to
the energy balance equation (\ref{2.6}).
In order to integrate the previous equations in a general spacetime, the
evolution equations for shear and vorticity are needed, together with
the dynamical equations for $G$ and $\Lambda$ which are obtained
by the RG equations.
The latter are obtained in the Einstein--Hilbert truncation as
a set of $\beta$-functions
for the dimensionless Newton constant and cosmological constant,
$g$ and $\lambda$,
\be\label{rg}
k{ \partial_k g } = \beta_g(g,\lambda), \;\;\;\;\;\;\;\;\;\;\;\;\;
k{ \partial_k \lambda } = \beta_{\lambda}(g,\lambda), \;\;\;\;\;\;
\ee
and the link with the spacetime dynamics is provided by
the  {\it cut-off identification}
\be\label{ide1}
k=k(\tau,\rho,\dot{\rho},\Theta, \dot{\Theta},...).
\ee
The dots stand for all possible physical or geometrical invariants
which can act as IR regulators in the fluctuation determinant of $\Gamma_k$.
The knowledge of the precise functional dependence in Eq. (\ref{ide1})
would then provide a dynamical evolution which is consistent
with the full effective action at $k=0$. In Ref.
\cite{2002PhRvD..65d3508B} the simple choice $k\propto 1/t$ can be justified on the
ground that, if there are no other scales in the system when the Universe had
age $t$, fluctuations with frequency greater than $1/t$
may not have played any role as yet, and the running must be stopped at
$k\propto 1/t$. On the other hand, recent works 
\cite{2004ForPh..52..650R,2005JCAP...09..012R,2007JCAP...08..024}
have argued that the Hubble parameter defined in
Eq.(\ref{hubble}) is a physically meaningful cutoff as it measures the 
curvature of the spacetime, and it also reproduces the $1/t$ cutoff for
any power-law dependence of $\ell(t)$ in Eq.(\ref{lun}). 
In fact the scalar curvature $R$, the square of the
Ricci tensor ${\cal R}= R_{\alpha\beta}R^{\alpha\beta}$ and the
Kretschmann invariant $K=R_{\alpha\beta\gamma\delta}R^{\alpha\beta\gamma\delta}$ can all be
expressed in terms of the Hubble parameter and its derivatives, 
\be\label{link}
R=6(2H^2+\dot H), \;\;\;\; {\cal R}=12(3H^4+3H^2\dot H+\dot H^2),\;\;\; 
K=12(2H^4 +2H^2 \dot H+\dot H ^2)
\ee
On the other hand, the second functional  derivative of the effective average
action reads \cite{2005JHEP...02..035B} 
${\Gamma}^{(2)}_k{[g,g]^{\mu\nu}}_{\rho\sigma} = 2\kappa^2 Z_{Nk}
[-{ K^{\mu\nu}}_{\rho\sigma}D^2+{U^{\mu\nu}}_{\rho\sigma}]$
%\ee
where\\
\be\label{11}
{ K^{\mu\nu}}_{\rho\sigma}= \frac{1}{4}[\delta^\mu_\rho\delta^\nu_\sigma+\delta^\mu_\sigma\delta^\nu_\rho-g^{\mu\nu}g_{\rho\sigma}]
\ee
and
\ba\label{12}
&&{U^{\mu\nu}}_{\rho\sigma} =
\frac{1}{4}[\delta^\mu_\rho\delta^\nu_\sigma+\delta^\mu_\sigma\delta^\nu_\rho-g^{\mu\nu}g_{\rho\sigma}](R-2\bar{\lambda}_k)
+ \frac{1}{2}[ g^{\mu\nu}R_{\rho\sigma}+g_{\rho\sigma}R^{\mu\nu}]\nonumber\\[2mm]
&&-\frac{1}{4}[\delta^\mu_\rho {R^{\nu}}_\sigma+\delta^\mu_\sigma {R^{\nu}}_\rho+\delta^\nu_\rho {R^{\mu}}_\sigma
+\delta^\nu_\sigma {R^{\mu}}_\rho]
-\frac{1}{2}[{{{R^{\nu}}_\rho}^\mu}_\sigma + {{{R^\nu}_\sigma}^\mu}_\rho]
\ea
which clearly shows the Ricci scalar, the Ricci tensor and the Riemann tensor
enters in the fluctuation determinant as mass-type regulators. 
As from (\ref{link}) all those terms are essentially
expressible in terms of the Hubble parameter, it is then clear that one can
conveniently parameterize the field strength dependence in terms of the single
scalar $H$. 
It is important to stress that this is a consequence of the very
high  degree of symmetry of our spacetime, but in a generic spacetime the
actual cutoff can be different.

\section{The improved Robertson-Walker cosmology}
Let us now  specify our spacetime 
to describe a spatially flat $(K=0)$ Robertson-Walker metric with scale factor
$\ell(t)$, so that the shear, rotation and acceleration are identically
vanishing.
We can take ${T_\mu}^\nu = {\rm diag}[-\rho,p,p,p]$ to be the energy
momentum tensor of an ideal fluid with equation of state $p=w\rho$ where $w>-1$ is constant.

Then the improved Einstein equation boils down 
to the modified Friedmann equation and a continuity equation:
\ba\label{zatt}
&&H^2 = \frac{8\pi}{3} G(t) \; \rho + \frac{1}{3}\Lambda(t)\\[2mm]
&&\dot\rho+3H(\rho+p)=-\frac{\dot{\Lambda}+8\pi \; \rho \; \dot{G}}{8\pi \;
G}\label{zb}
\ea
The modified continuity equation above is the integrability condition
for the improved Einstein equation implied by Bianchi's identity, 
$D^\mu [-\Lambda(t) g_{\mu\nu} + 8\pi G(t) T_{\mu\nu}]=0$. 
It describes the energy exchange between the matter and gravitational degrees of freedom (geometry).
For later use let us note that upon defining the critical density
$\rho_{\rm crit}(t)\equiv {3 \; H(t)^2}/{8\pi \; G(t)}$
and the relative densities $\Omega_{\rm M}\equiv \rho/\rho_{\rm crit}$ and 
$\Omega_\Lambda=\rho_\Lambda/\rho_{\rm crit}$ the modified Friedmann equation
(\ref{zatt}) can be written as
$\OM(t) +\OL(t) = 1$.

It is possible to obtain $G(k)$ and $\Lambda(k)$ by solving the flow equation in
the Einstein-Hilbert truncation with a sharp cutoff \cite{1998PhRvD..57..971R,2005JCAP...09..012R}. It is formulated in
terms of the dimensionless Newton and cosmological constant, respectively:
$g(k)\equiv k^2 \; G(k)$, $\lambda(k) = \Lambda(k)/k^2$.
Quantum corrected cosmologies are computed by (numerically) solving the RG
improved evolution equations. The cutoff identification
\be\label{2.12}
k(t) =\xi H (t)
\ee
where $\xi$ is a fixed positive constant of order unity will be employed.
As discussed in the previous section this is a natural choice since in a
Robertson-Walker geometry the Hubble parameter measures the curvature of
spacetime which is related to the actual regulator.
Thus we have
\be\label{2.13}
G(t) = \frac{g(\xi H(t))}{\xi^2 \; H(t)^2}, 
\;\;\;\;\;\;\;\;\;\; \Lambda(t) = \xi^2 \; H(t)^2 \; \lambda(\xi H(t))
\ee

Let us briefly review how the type IIIa trajectories of the Einstein-Hilbert truncation
can be matched against the observational data. This analysis is fairly robust 
and clearcut; it does not involve the NGFP. All that is needed is the RG flow linearized 
about the Gaussian fixed point (GFP) which is located at $g=\lambda=0$. In its vicinity one
has \cite{2007JCAP...08..024B}
$\Lambda(k) = \Lambda_0 + \nu \; \barg \, k^4+\cdots$ and $G(k) =\barg+\cdots$.
Or, in terms of the dimensionless couplings, 
$\lambda(k) = \Lambda_0/k^2 + \nu \; \barg \, k^2+\cdots$, $g(k) =\barg \; k^2+\cdots$.
In the linear  regime of the GFP, $\Lambda$ displays a running $\propto k^4$
and $G$ is approximately constant. Here $\nu$ is a positive 
constant of order unity \cite{2005JCAP...09..012R},  
$\nu \equiv \frac{1}{4\pi} \Phi^{1}_{2}(0)$. These equations are valid if $\lambda(k) \ll 1$ and $g(k)\ll 1$. They
describe a 2-parameter family of RG trajectories labeled by the pair $(\Lambda_0, \barg)$.
It will prove convenient to use an alternative labeling $(\lat, \kat)$ with
$\lat \equiv (4  \nu  \Lambda_0 \barg)^{1/2}$ and 
$\kat \equiv   ( {\Lambda_0}/{\nu  \barg}  )^{1/4} $.
The old labels are expressed in terms of the new ones as
$\Lambda_0 = \frac{1}{2} \lat \; \kat^2$ and $\barg = {\lat}/{2\,\nu\,\kat^2}$.
It is furthermore convenient to introduce the abbreviation
$\gat\equiv {\lat}/{2\,\nu}$.

When parameterized by the pair $(\lat,\kat)$ the trajectories assume the form
\ba\label{1.16}
&&\Lambda(k) = \frac{1}{2} \; \lat\; \kat^2 \; \Big [ 1+(k/\kat)^4\Big ]
\equiv \Lambda_0 \Big [ 1+(k/\kat)^4 \Big ]\\[2mm]
&&G(k) = \frac{\lat}{2\,\nu\,\kat^2}\equiv \frac{\gat}{\kat^2}\nonumber
\ea
or, in dimensionless form, 
\be\label{1.17}
\lambda(k) = \frac{1}{2} \; \lat \Big  [  \Big (\frac{\kat}{k} \Big)^2+ 
 \Big ( \frac{k}{\kat}  \Big)^2 \Big ], \;\;\;\;\;\;\;\;\;\;  g(k) = \gat \,  \Big( \frac{k}{\kat}\Big  )^2
\ee
As for the interpretation of the new variables, 
it is clear that $\lat \equiv \lambda(k\equiv \kat)$ and $\gat\equiv g(k=\kat)$,
while $\kat$ is the scale at which $\beta_\lambda$ (but not $\beta_g$) vanishes according
to the linearized running:
$\beta_\lambda(\kat)\equiv k{d \lambda(k)}/{dk}  |_{k=\kat} =0$.
Thus we see that $(\gat,\lat)$ are the coordinates of the turning point T
of the type IIIa trajectory considered, and $\kat$ is the scale at which it is
passed. It is convenient to refer the ``RG time'' $\tau$ to this scale:
$\tau(k) \equiv \ln (k/\kat)$.
Hence $\tau>0$ ($\tau<0$) corresponds to the ``UV regime'' (``IR regime'') where
$k>\kat$ ($k<\kat)$. 

Let us now hypothesize that, within a certain range of $k$-values, the RG trajectory
realized in Nature can be approximated by (\ref{1.17}). In order to determine
its parameters $(\Lambda_0, \barg)$ or $(\lat, \kat)$ we must perform a measurement
of $G$ and $\Lambda$. If we interpret the observed values
%\ba\label{1.21}
$G_{\rm observed} = \mp^{-2}$, $\mp\approx 1.2\times 10^{19} \, {\rm GeV}$, and 
$\Lambda_{\rm observed} = 3\,\Omega_{\Lambda 0}\,H_0^2\approx 10^{-120}\, \mp^2\nonumber$
as the running $G(k)$ and $\Lambda(k)$ evaluated at a scale $k\ll \kat$, then we get from
(\ref{1.16}) that $\Lambda_0 =\Lambda_{\rm observed} $ and 
$\barg = G_{\rm observed}$. Using the definitions of  $\lat$ and $\kat$ along with $\nu = O(1)$ this leads to the 
order-of-magnitude estimates
$\gat\approx \lat \approx 10^{-60}$ and $\kat\approx 10^{-30}\;\mp\approx (10^{-3} {\rm cm})^{-1}$.
Because of the tiny values of $\gat$ and $\lat$ the turning point lies in the linear regime of the GFP. 

Up to this point we discussed only that segment of the ``trajectory realized in Nature'' which lies
inside the linear regime of the GFP. The complete RG trajectory is obtained by
continuing this segment with the flow equation both into the IR and into the UV, 
where it ultimately spirals into the 
NGFP. While the UV-continuation is possible within the Einstein-Hilbert
truncation, this approximation breaks down in the IR when $\lambda(k)$ approaches $1/2$.
Interestingly enough, this happens near $k=H_0$, the present Hubble scale.
The right panel of Fig.1 shows a schematic sketch of the complete trajectory 
on the $g$-$\lambda$--plane and Fig.2 displays the
resulting $k$-dependence of $G$ and $\Lambda$.
%%%%%%%%%%%%%%%%%%%%%%%%%%%%%%%%%%%%%%%%%%%%%%%%%%5
\section{Primordial entropy generation}
%%%%%%%%%%%%%%%%%%%%%%%%%%%%%%%%%%%%%%%%%%%%%%%%
Let us return to the modified continuity equation (\ref{2.6}). After
multiplication by $a^3$ it reads
\be\label{3.1}
[\dot\rho + 3H(\rho +p)] \; a^3 = \calp(t)
\ee
where we defined
\be\label{3.2}
\calp\equiv -\Big ( \frac{\dot{\Lambda}+8 \pi\; \rho \; \dot{G}}{8\pi \; G} \Big ) a^3
\ee
Without assuming any particular equation of state eq.(\ref{3.1}) can be
rewritten as 
\be\label{3.3}
\frac{d}{dt} (\rho a^3) +p\frac{d}{dt}(a^3) = \calp(t)
\ee
The interpretation of this equation is as follows. 
Let us consider a unit {\it coordinate}, 
i.e. comoving volume in the Robertson-Walker spacetime. Its corresponding  
{\it proper} volume is $V=a^3$
and its energy contents is $U=\rho a^3$. The rate of change of these quantities is subject to 
(\ref{3.3}): 
\be\label{3.4}
\frac{dU}{dt}+p\frac{dV}{dt}=\calp(t)
\ee
In classical cosmology where $\calp\equiv 0$ this equation together with the standard thermodynamic
relation $dU+pdV=TdS$ is used to conclude that the expansion of the Universe is adiabatic, \ie
the entropy inside a comoving volume does not change as the Universe expands, $dS/dt=0$.

Here and in the following we write $S\equiv s \, a^3$ for the entropy carried by the matter inside
a unit comoving volume and $s$ for the corresponding proper entropy density.

When $\Lambda$ and $G$ are time dependent, $\calp$ is nonzero and we interpret (\ref{3.4})
as describing the process of energy (or ``heat'') exchange between the scalar fields $\Lambda$ 
and $G$  and the ordinary matter. This interaction causes $S$ to change:
\be\label{3.5}
T\frac{dS}{dt}=T\frac{d}{dt}(s a^3)=\calp(t)
\ee
The actual rate of change of the comoving entropy is 
\be\label{3.6}
\frac{dS}{dt}=\frac{d}{dt}(s a^3)= \clp (t)
\ee
where 
\be\label{3.7}
\clp \equiv \calp /T
\ee 
If $T$ is known as a function of $t$ we can integrate (\ref{3.5}) to obtain $S=S(t)$. 
In the RG improved cosmologies the entropy production rate per comoving 
volume 
\be\label{3.7}
\clp(t) = - \Big [ \frac{ \dot\Lambda+8\pi \; \rho \; \dot G}{8\pi\;  G} \Big ] \frac{a^3}{T}
\ee
is nonzero because the gravitational ``constants''  $\Lambda$ and $G$ have acquired a time 
dependence. 

Clearly we can convert the heat exchanged, $TdS$, to an entropy change only if the dependence
of the temperature $T$ on the other thermodynamical quantities, in particular $\rho$ and $p$
is known.  For this reason we shall now make the following assumption about the matter system and its
(non-equilibrium!)  thermodynamics:

{\it The matter system is assumed to consist 
of $n_{\rm eff}=n_{\rm b}+\frac{7}{8}n_{\rm f}$ species of effectively massless degrees of freedom
which all have the same temperature $T$. The equation of state is $p=\rho/3$, 
\ie $w=1/3$, and $\rho$ depends on $T$ as 
\be\label{3.9}
\rho(T) =\kappa^4 \; T^4 , \;\;\;\;\;\;\; \kappa\equiv (\pi^2 \; n_{\rm eff}/30)^{1/4}
\ee
No assumption is made about the relation $s=s(T)$.}

The first assumption, radiation dominance and equal temperature, is plausible since we shall find
that there is no significant entropy production any more once $H(t)$ has dropped substantially below
$\mp$.
The second assumption, eq.(\ref{3.9}), refer to the hypothesis that the
injection of energy into the matter system disturbs its equilibrium only very weakly. 
The approximation is that the 
{\it equilibrium} relations among $\rho$, $p$, and $T$ are still valid in the
non-equilibrium situation of a cosmology with entropy production. 

By inserting $p=\rho/3$ and (\ref{3.9}) into the modified continuity equation  
the entropy production rate can be seen to be
a total time derivative:
$\clp(t) = \frac{d}{dt}[ \frac{4}{3}  \kappa  a^3  \rho^{3/4}  ] $.
Therefore we can immediately integrate (\ref{3.6}) and obtain 
%\be\label{3.22}
$S(t)=\frac{4}{3} \kappa  a^3 \rho^{3/4} +S_{\rm c}$, 
%\ee
%or, in terms of the proper entropy density, 
%\be\label{3.23}
%s(t) = \frac{4}{3} \kappa \rho^{3/4}(t) +\frac{S_{\rm c}}{a^3(t)}
$s(t) = \frac{4}{3}  \kappa \rho(t)^{3/4} +\frac{S_{\rm c}}{a(t)^3}$.
%\ee
Here $\ssc$ is a constant of integration. In terms of $T$, using 
(\ref{3.9}) again, 
\be\label{3.24}
s(t) = \frac{2\pi^2 }{45} \; n_{\rm eff} \; T(t)^3 +\frac{S_{\rm c}}{a(t)^3}
\ee

The final result (\ref{3.24}) is very remarkable for at least two reasons.
Firstly, for $\ssc=0$, eq.(\ref{3.24}) has exactly the form  valid for radiation
{\it in equilibrium}. Note that we did not postulate this relationship, only the $\rho(T)$--law
was assumed. The equilibrium formula $s\propto T^3$ was {\it derived} from the 
cosmological equations, \ie the modified conservation law. This result makes the hypothesis
``non-adiabatic, but as little as possible'' selfconsistent. 

Secondly, if $\lim_{t\rightarrow 0} \; a(t) \rho(t)^{1/4}=0$, which is  actually
the case for the most interesting class of cosmologies we shall find, then $S(t\rightarrow 0)=S_c$.
As we mentioned in the introduction, the most plausible initial value
of $S$ is $S=0$ which means a vanishing constant of integration $S_c$ here. But then,
with $S_c=0$ the {\it entire} entropy carried by the 
massless degrees of freedom is due to the RG running. So it indeed seems to 
be true that the entropy of the CMBR photons we observe today is due to a
coarse graining. 
Unexpectedly, not a coarse graining of the matter degrees of freedom but rather
of the gravitational ones which determines the background spacetime the photons
propagate on.
%%%%%%%%%%%%%%%%%%%%%%%%%%%%%%%%%%%%%%%%%%%%%

\section{Solving the RG improved Einstein Equations}
In \cite{2007JCAP...08..024B} the improved Einstein equations (\ref{zatt},
\ref{zb}) have been solved  for the trajectory with realistic parameter values
which was discussed in Section 3. The solutions were determined by applying 
the algorithm described at the end of Section 2. Having fixed the RG trajectory, there exists 
a 1-parameter family of solutions $(H(t),\rho(t))$. This parameter is conveniently chosen to be the relative
vacuum energy density in the fixed point regime, $\OLS$. 

The very early part of the cosmology can be described analytically. For $k\rightarrow \infty$ the trajectory
approaches the NGFP, $(g,\lambda)\rightarrow (g_\ast,\lambda_\ast)$,  so that $G(k)=g_\ast/k^2$ and 
$\Lambda(k)=\lambda_\ast k^2$. In this case the differential equation can be solved analytically, with the result
\be\label{4.12}
H(t)=\alpha/t, \;\;\;\; a(t) = At^\alpha, \;\;\;\;\; \alpha = \Big [ \frac{1}{2}(3+3w)(1-\OLS) \Big]^{-1}
\ee
and  $\rho(t)=\widehat\rho t^{-4}$, $G(t) = \widehat G  t^2$, $\Lambda(t) = \widehat\Lambda  /t^2$.
Here $A$, $\widehat\rho$, $\widehat G $, and $\widehat\Lambda $ are positive constants.
They depend on $\OLS$ which assumes
values in the interval $(0,1)$.

Summarizing the numerical results one can say  that for any value of $\OLS$ the UV cosmologies
consist of two scaling regimes and a relatively sharp crossover region near
$k,H\approx \mp$ corresponding to $x\approx -34.5$  
which connects them. At higher $k$-scales the fixed point approximation 
is valid, at lower scales one has a classical FRW cosmology in which $\Lambda$
can be neglected.

\begin{figure}[h]
\begin{center}
\includegraphics[width=6in]{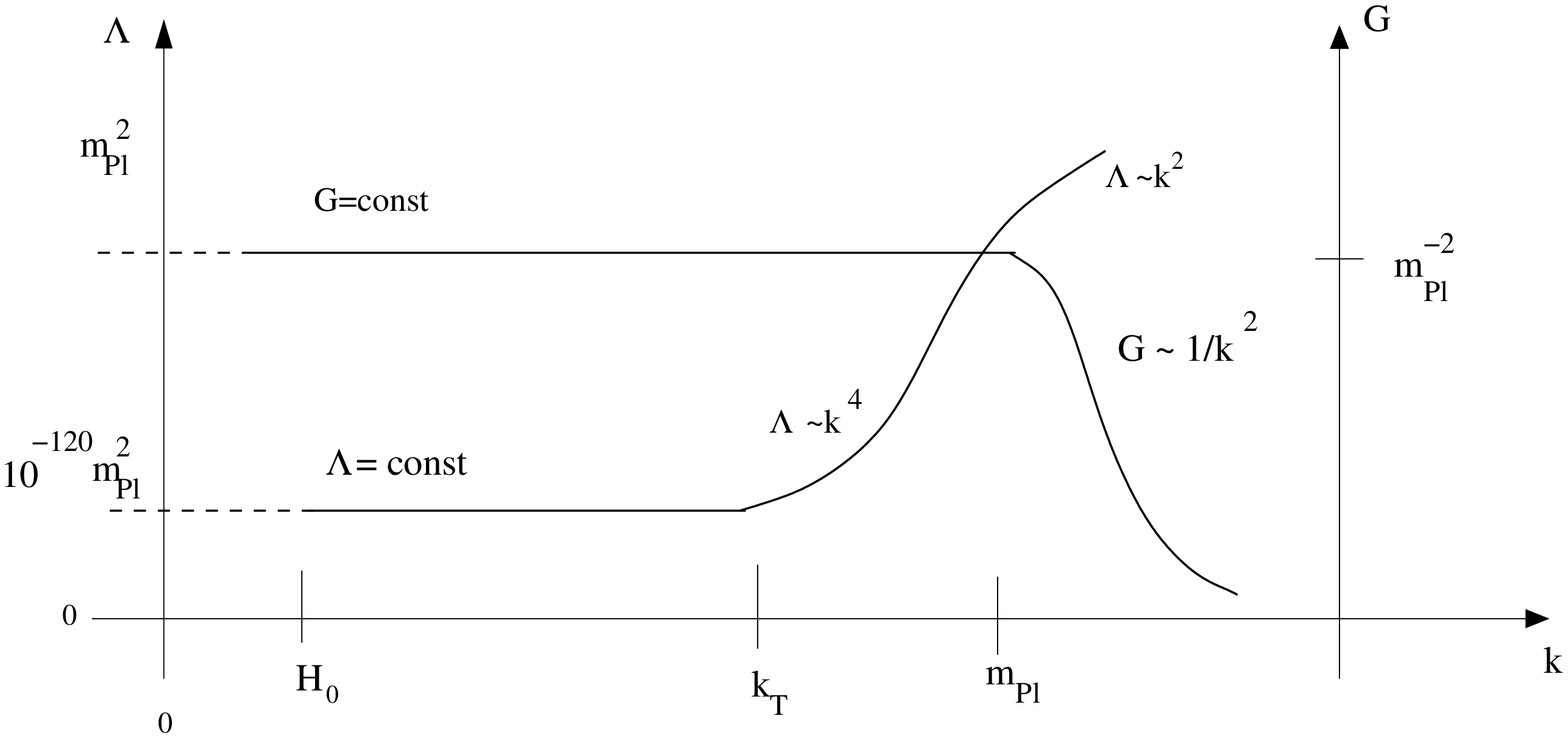}
\end{center}
\caption{The dimensionful quantities $\Lambda(k)$ and $G(k)$ for the RG trajectory with
realistic parameter values.}
\end{figure}

\begin{figure}[t]
\begin{center}
\includegraphics[width=15cm, height=11cm]{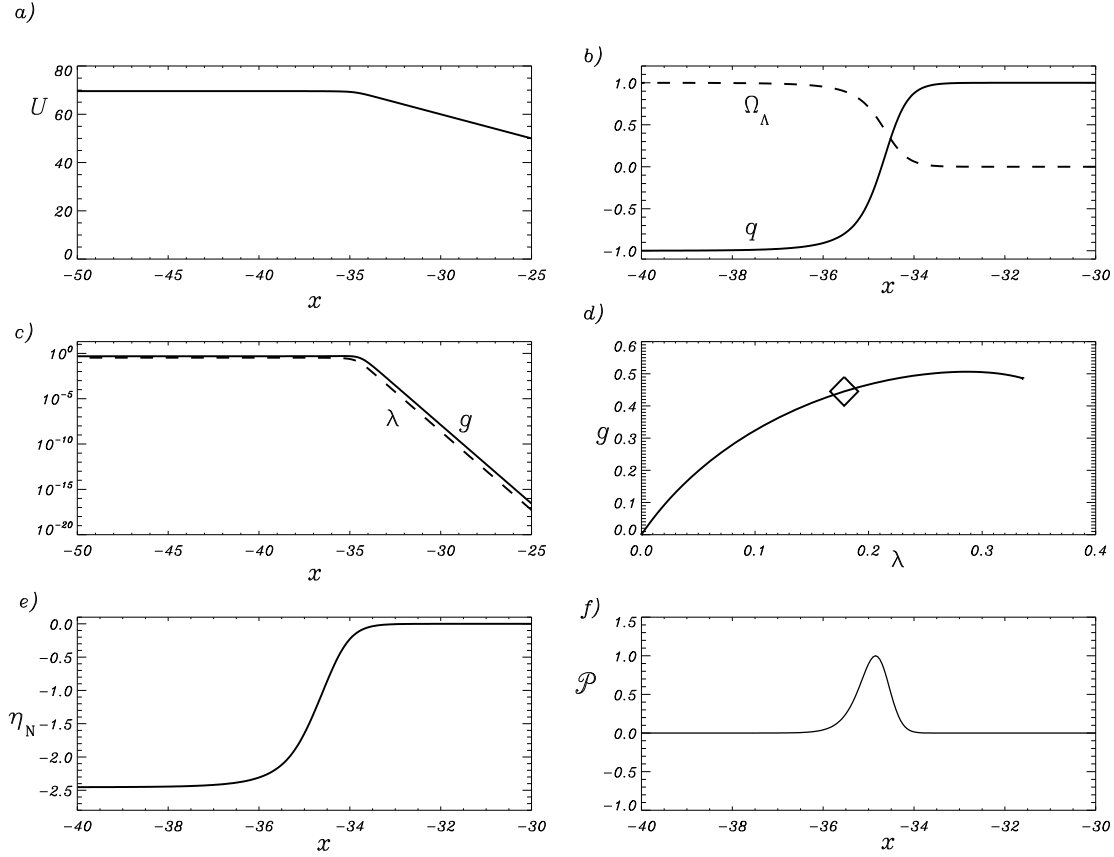}
\end{center}
\caption{The  crossover epoch of the cosmology for $\OLS=0.98$. The plots a), b), c) display the logarithmic Hubble
parameter $\UU$, as well as $q$, $\OL$, $g$ and $\lambda$ as a function of the logarithmic
scale factor $x$. A crossover is observed near $x\approx -34.5$. 
The diamond in plot d) indicates the point on the RG trajectory corresponding to this $x$-value. 
(The lower horizontal part of the trajectory is not visible on this scale.)
The plots e) and f) show the $x$-dependence of the anomalous dimension and entropy
production rate, respectively.}
\label{fig7}
\end{figure}
As an example, Fig.(\ref{fig7}) shows the crossover cosmology with  $\OLS=0.98$ and $w=1/3$. 
The entropy production rate $\clp$ is maximum at $t_{\rm tr}$ and quickly goes to zero for 
$t>t_{\rm tr}$; it is non-zero for all $t<t_{\rm tr}$. By varying the $\OLS$-value one can
check that the early cosmology is indeed described by the NGFP solution (5.1). For the logarithmic
$H$ vs. $a$- plot, for instance, it predicts $\UU=-2(1-\OLS)x $ for $ x<-34.4$. The left part of the plot in
Fig.3a and its counterparts with different values of $\OLS$ indeed comply with this relation.
If $\OLS \in (1/2,1)$ we have $\alpha = (2-2\OLS)^{-1}>1$ and $a(t)\propto t^\alpha$ describes a phase of
accelerated power law inflation. 

%When $\OLS \nearrow 1$ the slope of $\UU (x) = -2 (1-\OLS)x$ decreases and
%finally vanishes at $\OLS=1$. This limiting case corresponds  to a constant Hubble parameter, 
%i.e. to de Sitter space. For values of $\OLS$ smaller than, but close to $1$ this 
%de Sitter limit is approximated by an expansion $a\propto t^\alpha$ with a very large
%exponent $\alpha$. 
The phase of power law inflation automatically comes to a halt 
once the RG running has reduced $\Lambda$ to a value where the resulting vacuum energy density no longer
can overwhelm the matter energy density.

\section{Inflation in the fixed point regime}
%%%%%%%%%%%%%%%%%%%%%%%%%%%%%%%%%%%%%%%%%%%%%%%%%%%%%%%%%%%%%%%%%%%%%%%%%%%
Next we discuss in more detail the epoch of power law inflation which is realized
in the NGFP regime if $\OLS > 1/2$. Since 
the transition from the fixed point to the classical FRW regime is rather sharp it will be
sufficient to approximate the RG improved UV cosmologies by the following caricature :
For $0<t< t_{\rm tr}$, the scale factor behaves as
$a(t)\propto t^\alpha$, $\alpha > 1$.  Here $\alpha = (2-2\OLS)^{-1}$ since $w =1/3$ 
will be assumed. Thereafter, for $t>t_{\rm tr}$, we have a classical, entirely
matter-driven expansion $a(t)\propto t^{1/2}$ . 
%%%%%%%%%%%%%%%%%%%%%%%%%%%%%%%%%%%%%%%%%%%%%%%%%%%%%%%%%%%%%%%%%%%%%%%%%%%%%%%%%%
%\subsection{Transition time and apparent initial singularity}

The transition time $t_{\rm tr}$ is dictated by the RG trajectory. 
It leaves the asymptotic scaling 
regime near $k\approx \mp$. Hence  $H(t_{\rm tr})\approx \mp$ and since $\xi=O(1)$ and $H(t)=\alpha/t$
we find the estimate
\be\label{6.2}
t_{\rm tr}= \alpha \; \tp
\ee
Here, as always, the Planck mass, time, and length are defined in terms of the value of Newton's
constant in the classical regime :
$\tp = \lp = \mp^{-1} = \bar{G}^{1/2} = G_{\rm observed}^{1/2}$.
Let us now assume that $\OLS$ is very close 
to $1$ so that $\alpha$ is large:
$\alpha \gg 1$. Then (\ref{6.2}) implies that the transition takes place at a cosmological time 
which is much later 
than the Planck time. At the transition the {\it Hubble parameter} is of order $\mp$, but the 
{\it cosmological time } is in general not of the order of $\tp$. Stated differently, the ``Planck time''
is {\it not} the time at which $H$ and the related physical quantities assume Planckian values. 
The Planck time as defined above is well within the NGFP regime:
$\tp = t_{\rm tr} / \alpha \ll t_{\rm tr}$. 

At $t=t_{\rm tr}$ the NGFP solution is to be matched continuously with a FRW cosmology
(with vanishing cosmological constant ). We may use the  classical formula $a\propto \sqrt{t}$ 
for the scale factor, but we must shift the time axis on the classical side such that $a$, 
$H$, and then as a result of (\ref{zatt}) also $\rho$ are continuous at $t_{\rm
tr}$. Therefore $a(t)\propto (t-t_{\rm as})^{1/2}$ and 
$H(t) = \frac{1}{2} \; (t-t_{\rm as})^{-1} \;\;\; \text{for } \;\;\; t> t_{\rm tr}$.
Equating this Hubble parameter  at $t=t_{\rm tr}$ to 
$H(t) = \alpha /t$, valid in the NGFP regime, we find that the shift $t_{\rm as}$ must be chosen
as $t_{\rm as} =  (\alpha -\frac{1}{2}) \tp = (1 - \frac{1}{2\alpha})  t_{\rm tr}  <  t_{\rm tr}$. 
Here the subscript 'as' stands for ``apparent singularity''. This is to indicate that if one continues
the classical cosmology to times $t<t_{\rm tr}$, it has an initial singularity (``big bang'') at 
$t=t_{\rm as}$. Since, however, the FRW solution is not valid there nothing special happens at 
$t_{\rm as}$; the true initial singularity is located at $t=0$ in the NGFP regime. (See Fig.~4.)
%%%%%%%%%%%%%%%%%%%%%%%%%%%%%
\subsection{Crossing the Hubble radius}
In the NGFP regime $0<t< t_{\rm tr}$ the Hubble radius $\ell_{H} (t) \equiv 1/H(t)$, i.e.
$\ell_{H} (t) = t/{\alpha} $ , 
increases linearly with time but, for $\alpha \gg 1$, with a very small slope. At the transition, the slope 
jumps from $1/\alpha$ to the value $2$ since $H=1/(2t)$ 
and $\ell_{H}=2t$ in the  FRW regime. This behavior is sketched in Fig.~4. 

Let us consider some structure of comoving length $\Delta x$, 
a single wavelength of a density perturbation,
for instance. The corresponding physical, i.e. proper length is $L(t) = a(t) \Delta x$ then. 
In the NGFP regime it has the time dependence 
$L(t) =  ({t}/{\tr}  )^{\alpha} \; L(\tr)$.
The ratio of $L(t)$ and the Hubble radius evolves according to 
$\frac{L(t)}{\LH (t)} =  ( \frac{t}{\tr}  )^{\alpha-1} \; \frac{L(\tr)}{\LH (\tr)}$.
For $\alpha > 1$, i.e. $\OLS > 1/2$, the proper length of any object grows faster than the Hubble
radius. So objects which are of ``sub-Hubble'' size at early times can cross the Hubble radius and become
``super-Hubble'' at later times, see Fig.~4. 

Let us focus on a structure which, at $t=\tr$, is
$e^N$ times larger than the Hubble radius. Before the transition we have
$L(t)/\LH (t) = e^N \; (t/\tr)^{\alpha -1}$.
Assuming $e^N > 1$, there exists a time $t_N < \tr$ at which $L(t_{N}) =\LH (t_{ N})$
so that the structure considered ``crosses'' the Hubble radius at the time $t_N$. It is given
by 
\be\label{6.8}
%t_{N}=\tr \;{\rm exp}  {\Big ( -\frac{N}{\alpha -1} \Big )} = \tr \; {\rm exp} 
%\Big [ -\frac{(1-\OLS) N}{(\OLS-1/2)} \Big ]
t_{N}=\tr \;{\rm exp} {\Big ( -\frac{N}{\alpha -1} \Big )} 
\ee
What is remarkable about this result is that, even with rather moderate values of $\alpha$, one can 
easily ``inflate'' structures to a size which is by many $e$-folds larger than the Hubble radius 
{\it during a very short time interval at the end of the NGFP epoch}. 

Let us illustrate this phenomenon by means of an example, namely the choice $\OLS = 0.98$ used
in Fig.~3.
Corresponding to $98\%$ vacuum and $2\%$ matter energy density in the NGFP regime, this value
is  still ``generic'' in the sense that $\OLS$ is not fine tuned to equal unity 
with a precision of many decimal places. It leads to the exponent $\alpha = 25$, the transition
time $\tr = 25 \; \tp$, and $t_{\rm as}=24.5 \; \tp$. 

The largest structures in the present Universe, evolved backward in time by the classical equations
to the point where $H=\mp$, have a size of about $e^{60}\; \lp$ there. We can use 
(\ref{6.8}) with $N=60$ to find the time $t_{\rm 60}$ at which those structures crossed
the Hubble radius. With $\alpha = 25$ the result is $t_{\rm 60}=2.05\; \tp = \tr /12.2$. 
Remarkably, $t_{\rm 60}$ is smaller than 
$\tr$ by one order of magnitude only. As a consequence, the physical conditions prevailing at the
time of the crossing are not overly  ``exotic'' yet. 
The Hubble parameter, for instance, is only one order of magnitude larger than at the transition:
$H(t_{\rm 60})\approx 12 \mp$.  The same is true for the temperature; one can show that 
$T(t_{\rm 60})\approx 12 T(\tr)$ where $T(\tr)$ is of the order of $\mp$. Note  that $t_{\rm 60}$ is larger than $\tp$.

\subsection{Primordial density fluctuations}
%%%%%%%%%%%%%%%%%%%%%%%%%%%%%%%%%%%%%%%%%%%%%%%%%%%%%%%%%%%%%%%%
\begin{figure}[t]
\begin{center}\label{pep}
\includegraphics[width=10cm]{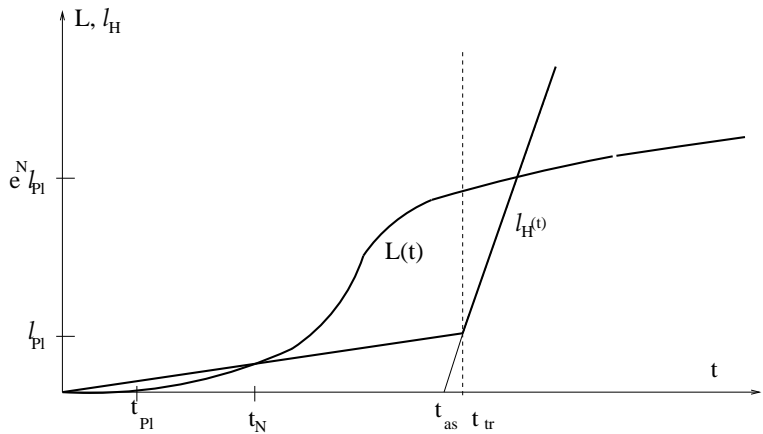}
\end{center}
\caption{Shown is the proper length $L$ and the Hubble radius as a function of time. 
The NGFP and FRW cosmologies are valid for $t<t_{\rm tr}$ and $t>t_{\rm tr}$, respectively.
The classical cosmology has an apparent initial singularity at $t_{as}$ outside its domain of 
validity. Structures of size $e^N \lp$ at $t_{\rm tr}$ cross the Hubble radius at $t_N$, a 
time which can be larger than the Planck time. }
\label{fig9}
\end{figure}

QEG offers a natural mechanism for generating primordial fluctuations during the NGFP epoch.  They  have a
scale free spectrum with a spectral index close to $n=1$. This mechanism is at the very heart of the 
``asymptotic safety'' underlying the non-perturbative renormalizability of QEG. 
A detailed discussion of this mechanism is beyond the scope of the present review; the reader it referred to 
\cite{2002PhRvD..65b5013L,2002PhRvD..66b5026L,2002PhRvD..65d3508B,2007JCAP...08..024B}. Suffice it
to say that the quantum mechanical generation of the primordial fluctuations happens on sub-Hubble distance scales. However, thanks to the inflationary NGFP era the modes relevant
to cosmological structure formation were indeed smaller than the Hubble radius at a sufficiently early time,
for $t<t_{\rm 60}$.

\section{RG improved Black Hole spacetimes}

In \cite{2000PhRvD..62d3008B}, a ``RG-improvement" of the Schwarzschild metric
has been performed and the properties of the corresponding ``quantum black hole" have been explored. 
The improvement was based upon the scale dependent (``running") Newton constant $G(k)$ obtained from the exact RG
equation for gravity describing the scale dependence of the effective
average action. 
In this case the effect of the cosmological constant has been neglected, and
the running of $G$ is approximately given by 
\be\label{2.23}
G(k)={G_0\over 1+\omega \; G_0 \; k^2}
\ee
where $G_0$ denotes the laboratory value of Newton's constant, and $\omega$ is a constant. At large distances 
$(k\rightarrow 0)$, $G(k)$ approaches $G_0$, and in the ultraviolet limit $(k\rightarrow \infty)$, 
it decreases as $G(k)\propto 1/k^2$. This is the fixed point behavior responsible for the conjectured 
non-perturbative renormalizability of Quantum Einstein Gravity, in the
approximation of neglecting the running of the Cosmological Constant.

In the RG improvement scheme of \cite{2000PhRvD..62d3008B} the information
about the $k$-dependence of $G$ is exploited in the following way. 
The starting point is the classical Schwarzschild 
metric 
\be\label{due}
ds^2 = -f(r) dt^2 + f(r)^{-1}dr^2 + r^2 d\Omega^2
\ee
with $d\Omega^2 \equiv d\theta^2 +\sin^2 \theta d\phi^2$ and the classical lapse function
$f(r)= 1-2G_0 M/r \equiv f_{\rm class}(r)$. The RG improvement is effected by substituting, 
in $f_{\rm class}(r)$, $G_0$ by the $r$-dependent Newton constant $G(r)\equiv G(k=k(r))$ which 
obtains from $G(k)$ via an appropriate ``cutoff identification" $k=k(r)$. In flat space 
the natural choice would be $k \propto 1/r$. In \cite{2000PhRvD..62d3008B}
it was argued that in the Schwarzschild background the correct choice,   
in leading order at least, is $k(r)= \xi/d(r)$ where $\xi$ is a constant of the 
order of unity, and $d(r)\equiv \int_0^r dr' | f_{\rm class}(r')|^{-1/2}$ is the proper distance from a point
with coordinate $r$ to the center of the black hole. 
While the integral defining $d(r)$ can be evaluated exactly, it is sufficient
to use the following approximation which becomes exact for both $r\rightarrow \infty$ and $r \rightarrow 0$: 
\be\label{tre}
d(r)=\left ( {r^3\over r+\gamma \; G_0\; M} \right )^{1\over 2}
\ee
The resulting $G(r)\equiv G(k=\xi/d(r))$ reads
\be\label{quattro}
G(r)={G_0 \; r^3\over r^3 +\tilde{\omega}\;G_0\; [r+\gamma G_0 M]}
\ee
where $\widetilde{\omega}\equiv \omega\xi^2$. In these equations the parameter $\gamma$ has the value
$\gamma =9/2$ if one sets $k=\xi/d(r)$ as above. It turns out, however, that most of the
qualitative properties of the improved metric, in particular all those related to the structure of its horizons,
are fairly insensitive to the precise value of $\gamma$. In particular, $\gamma=0$ (corresponding to $k=\xi/r$) 
and $\gamma =9/2$ where found \cite{2000PhRvD..62d3008B} to lead to rather similar results
throughout. For this reason one can  adopt the choice $\gamma=0$ in the present paper.
It has the advantage that with this choice many calculations can be performed analytically which require a numerical treatment otherwise.

The metric of the RG improved Schwarzschild black hole is given by the line element (\ref{due}) with 
\be\label{cinque}
f(r)=1-\frac{2 G(r) M}{r}
\ee
Let us briefly list its essential features
%\footnote{All formulas quoted refer to $\gamma=0$, but the 
%qualitative features are the same for $\gamma =9/2$; see
%\cite{2000PhRvD..62d3008B} for details.}.

a) There exists a critical mass value
%\footnote{We define the (standard) Planck mass and length 
%in terms of the laboratory value $G_0$: $m_{\rm Pl}=\ell_{\rm
%Pl}^{-1}=1/\sqrt{G_0}$.} 
\be\label{sei}
M_{\rm cr}=\sqrt{\widetilde{\omega}/G_0}=\sqrt{\widetilde{\omega}} \;  m_{\rm Pl}
\ee
such that $f(r)$ has two simple zeros at $r_{-}$ and $r_{+}>r_{-}$ if $M>M_{\rm cr}$, one double zero
at $r_{+}=r_{-} = \sqrt{\widetilde{\omega}G_0}$ if $M=M_{\rm cr}$, and no zero at all if $M<M_{\rm cr}$.
For $M>M_{\rm cr}$ the zeros are at 
\be\label{sette}
r_{\pm} = G_0 M \; [1\pm \sqrt{1-\Omega}]
\ee
with the convenient abbreviation 
\be\label{otto}
\Omega \equiv \frac{M_{\rm cr}^2 }{M^2} = \widetilde{\omega} \; \Big ( \frac{m_{\rm Pl}}{M} \Big )^2 
\ee
The spacetime has an outer horizon at $r_{+}$ and in inner (Cauchy) horizon at $r_{-}$ . At 
$M_{\rm cr}$, the black hole is extremal, the two horizons coincide, and
the spacetime is free from any horizon if the mass is sufficiently small,  $M<M_{\rm cr}$.

b) The Bekenstein-Hawking temperature $T_{\rm BH}= \kappa /2\pi$ is given by the surface gravity 
at the outer horizon, $\kappa = {1\over 2}f'(r_{+})$. Explicitly,
\be\label{nove}
T_{\rm BH}(M)  = {1\over 4\pi G_0 M}\;{\sqrt{1-\Omega}\over 1 +\sqrt{1-\Omega}}
={1\over 4\pi G_0 M_{\rm cr}} \; {\sqrt{\Omega(1-\Omega)} \over 1 +\sqrt{1-\Omega}}= 
\frac{M_{\rm cr}}{4\pi\widetilde{\omega}}
{\sqrt{\Omega(1-\Omega)}\over 1+\sqrt{1-\Omega}}
\ee
This temperature vanishes for $M \searrow M_{\rm cr}$, 
{\it i.e.} $\Omega \nearrow 1$, thus motivating the interpretation of the improved Schwarzschild 
metric with $M=M_{\rm cr}$ as describing a ``cold" remnant of the evaporation process.

c) The energy flux from the black hole, its luminosity $L$, can be estimated using Stefan's law. 
It is given by $L=\sigma {\cal A}(M) T_{\rm BH}(M)^4$ where $\sigma$ is a constant and ${\cal A}\equiv
4\pi r_{+}^2$ denotes the area of the outer horizon. With (\ref{sette}) and (\ref{nove})
we obtain
\be\label{dieci}
L(M) = {\sigma \; M_{\rm cr}^2\over (4\pi)^3 \; \widetilde{\omega}^2}\;
{\Omega (1-\Omega)^2\over [1+\sqrt{1-\Omega}]^2}
\ee 
For a single massless field with two degrees of freedom one has $\sigma = \pi^2 /60$.

\section{The quantum-corrected Vaidya metric}
%%%%%%%%%%%%%%%%%%%%%%%%%%%%%%%%%%%%%%%%%%%%%%%%%%%%%%%%%%%%%%%%%%%%%%%%%%%%%%%%%%%%%%
An important issue is to find a metric which describes the history of an
evaporating Schwarzschild black hole and its gravitational field
\cite{2006PhRvD..73h3005B}. In the small 
luminosity limit $(L\rightarrow 0)$ this metric is supposed 
to reduce to the static metric of the RG improved Schwarzschild spacetime.

By reexpressing the metric (\ref{due}) with the improved lapse function
(\ref{cinque}) in terms of ingoing Eddington-Finkelstein coordinates 
$(v,r,\theta,\phi)$ it is convenient to  trade the Schwarzschild time $t$
for the advanced time coordinate 
\begin{equation}\label{dueuno}
v=t+r^\star, \;\;\;\;\;\;\;\;\;\;\;\; r^\star \equiv \int^r dr' /f(r')
\end{equation}
Here $r^\star$ is a generalization of the familiar ``tortoise" radial coordinate to which it 
reduces if $G(r)=const$. For $G(r) \not = const$ the 
function $r^\star= r^\star(r)$ is more complicated, but its
explicit form will not be needed here. Eq.(\ref{dueuno}) implies 
$dv= dt+dr/f(r)$, turning (\ref{due}) with (\ref{cinque}) into 
\be\label{duedue}
ds^2=-[1-2G(r)M/r] \; dv^2 + 2 dv dr +r^2d\Omega^2
\ee
Eq.(\ref{duedue}) is exactly the Schwarzschild metric in Eddington-Finkelstein coordinates, with $G_0$
replaced by $G(r)$. It is thus reassuring to see that the two operations, the RG improvement 
$G_0\rightarrow G(r)$ and the change of the coordinate system, can be performed in either order, they 
``commute".

The thermodynamical properties derived in \cite{2000PhRvD..62d3008B} and summarized in the
previous section refer to the metric (\ref{duedue}). In the exterior of the hole the spacetime is static, 
and while we can deduce a temperature and a corresponding luminosity from its periodicity in imaginary time 
(or by computing the surface gravity  at  $r_{+}$ directly) the backreaction 
of the mass-loss due to the evaporation
is not described by (\ref{duedue}). From the static metric we obtained the mass dependence of the 
luminosity, $L=L(M)$. Using this information we can compute the mass of the hole as seen by 
a distant observer at time $v$, $M(v)$, by solving 
the differential equation
\be\label{duetre}
-{d \over dv} M(v) = L(M(v))
\ee
In our case $L(M)$ is given by Eq.(\ref{dieci}). 
To first order in the luminosity, the metric which incorporates the effect of the decreasing mass is obtained by 
replacing the constant $M$ in (\ref{duedue}) with the $M(v)$ obtained from Eq.(\ref{duetre}):
\be\label{duequattro}
ds^2=-[1-2G(r)M(v)/r] \; dv^2 + 2 dv dr +r^2d\Omega^2
\ee
For $G(r)=const$, Eq.(\ref{duequattro}) is the Vaidya metric which frequently had been used to explore 
the influence of the Hawking 
radiation on the geometry. It is a solution of Einstein's equation $G_{\mu\nu}=8\pi G_0 T_{\mu\nu}$
where $T_{\mu\nu}$ describes an inward moving null fluid. In this picture the decrease of $M$ is due to the
inflow of negative energy, as it is appropriate if the field whose quanta are radiated off 
is in the Unruh vacuum.

The metric (\ref{duequattro}) can be regarded as a RG improved Vaidya metric. 
It encapsulates two different mechanisms whose combined effect can be studied here: the black hole radiance, 
and the modifications of the spacetime structure due to the 
quantum gravity effects, the running of $G$ in particular. 

\begin{figure}
\begin{center}
\includegraphics[width=4.5cm]{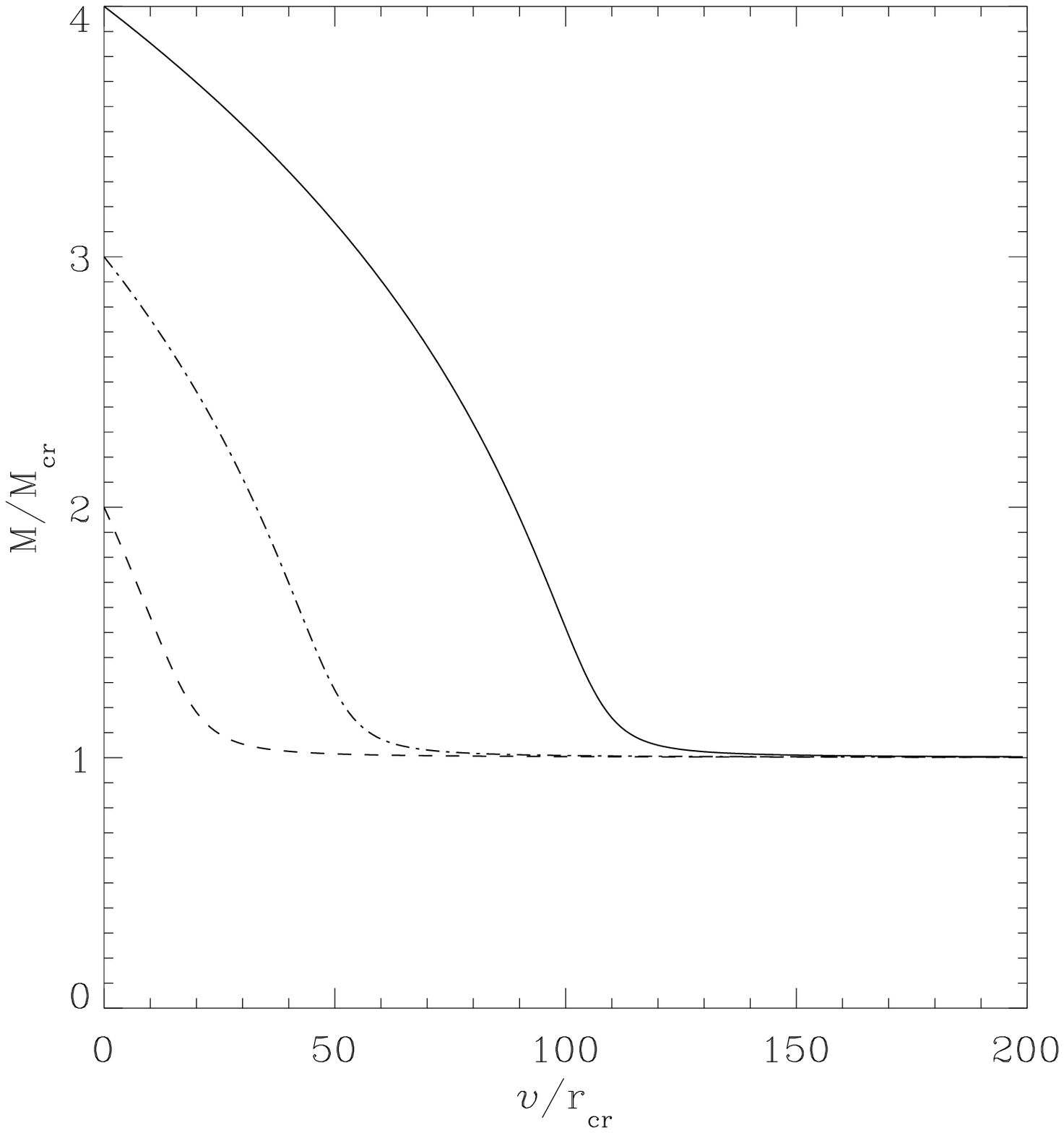}
\includegraphics[width=4.5cm]{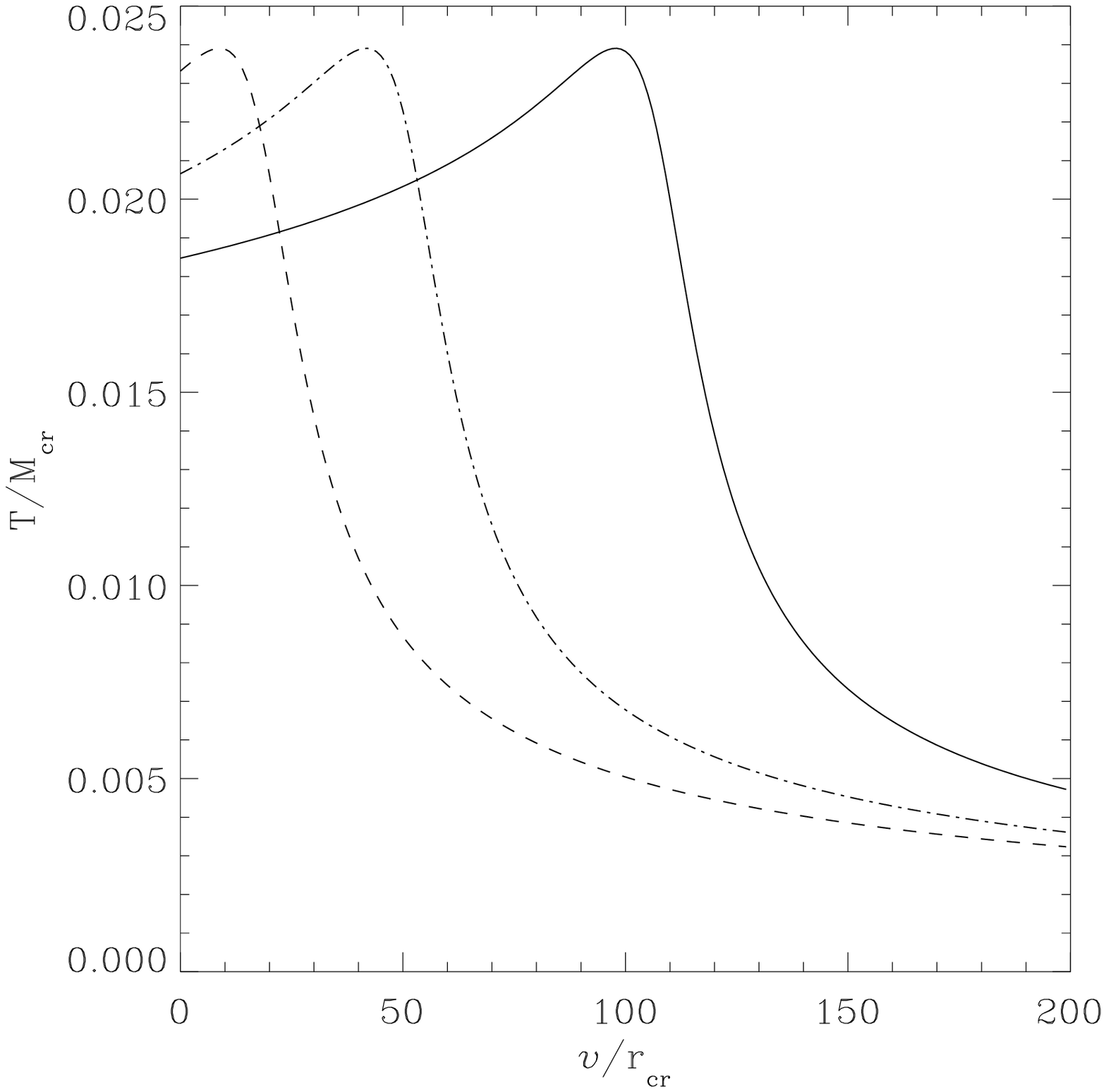}
\includegraphics[width=4.5cm]{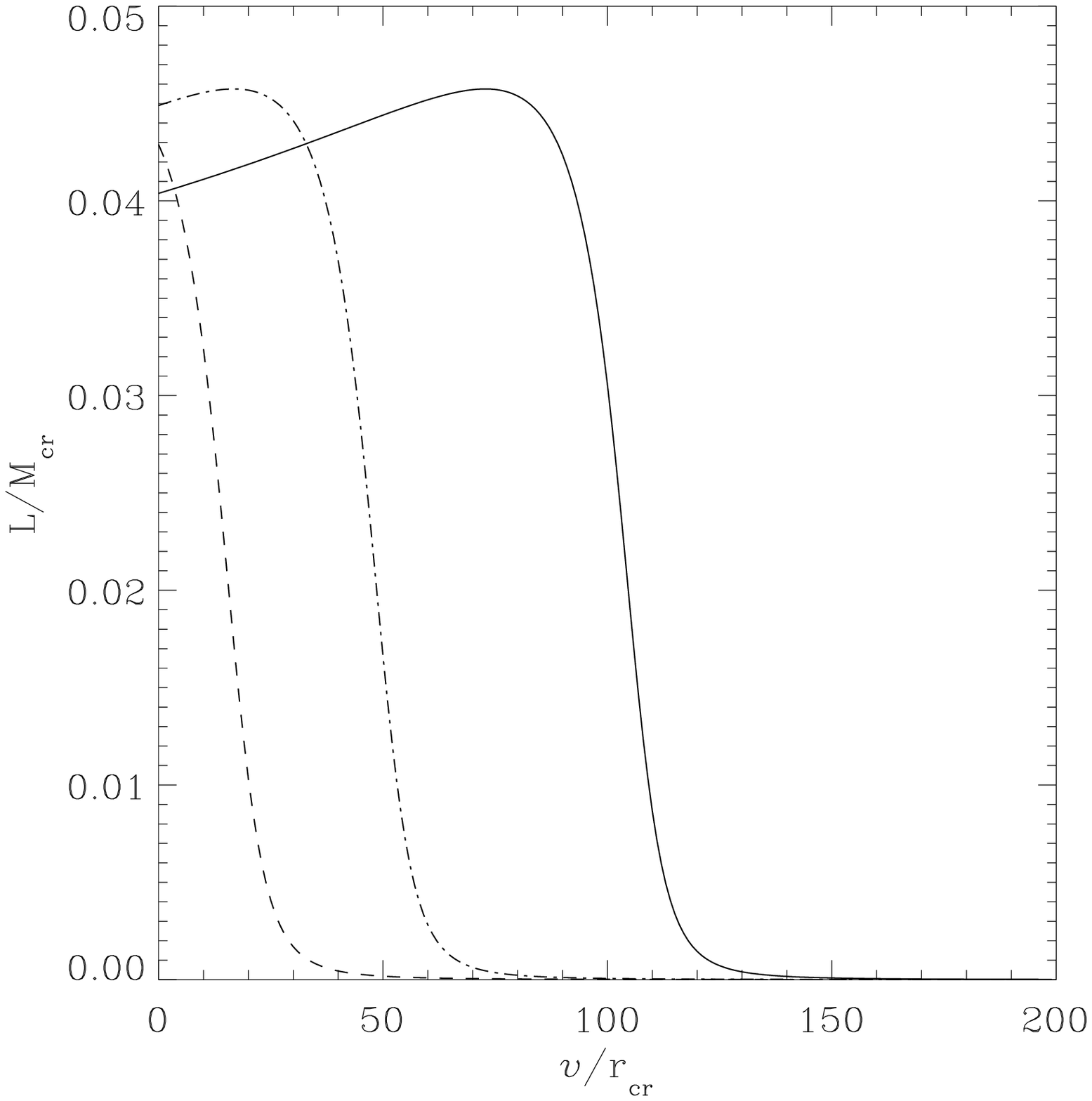}
\end{center}
\caption{\label{fig1}
The ratio $M/M_{\rm cr}$, the Bekenstein-Hawking temperature and the
BH luminosity as a function of $v/r_{\rm cr}$
%$(4\pi)^3\sqrt{\widetilde{\omega}}/\sigma m_{\rm Pl}$, for various initial masses. } 
for various initial masses, $M/M_{\rm cr}=1,2,3$, respectively. } 
\end{figure}

It is instructive to ask which energy-momentum tensor $T_{\mu\nu}$ would give rise to the improved
Vaidya metric (\ref{duequattro}) according to the classical equation ${G_{\mu}}^{\nu}=8\pi G_0 {T_{\mu}}^{\nu}$. Computing the 
Einstein tensor of (\ref{duequattro}) one finds that its only non-zero components are
\begin{subequations}\label{2.5}
\ba
&& {T^{v}}_v = {T^r}_r=-\frac{G'(r)M(v)}{8\pi G_0 r^2}\label{2.5a}\\[2mm]
&& {T^r}_v = \frac{G(r) \dot{M}(v)}{8\pi G_0 r^2}\label{2.5b}\\[2mm]
&& {T^{\theta}}_\theta =  {T^\phi}_\phi= -\frac{G''(r) M(v)}{16\pi G_0 r} 
\ea
\end{subequations}
Here the prime (dot) denotes a derivative with respect to $r(v)$. The non-zero components 
(\ref{2.5}) contain either $r$- or 
$v$-derivatives but no mixed terms. The terms with $r$-derivatives of $G$, also present for $M(v)=const$, describe the
vacuum energy density and pressure of the improved Schwarzschild spacetime in absence of radiation effects. 
Allowing for $M(v)\not = const$, the new feature is a nonzero component ${T^r}_v\not = 0$
which, for $\dot {M}< 0$, describes the inflow of negative energy into the black hole.

Taking advantage of the luminosity function $L(M)$, 
Eq.(\ref{dieci}), we can solve the differential equation (\ref{duetre})  
numerically and obtain the mass function $M=M(v)$. (We have 
set $\sigma / (4\pi)^3 \widetilde{\omega} = 1$ in the 
numerical calculations in order to reach the almost complete evaporation  
for $v\approx 200$ in units of $r_{\rm cr}$.) 
%%%%%%%%%%%%%%%%%%%%%%%%%%%%%% FIG 1
The result is shown in Fig.(\ref{fig1}) for various initial masses, in the domain $v>0$. In fact, 
for definiteness we assume that the black hole is formed at $v=0$ by the implosion of a spherical
null shell. Hence $M(v)$ is given by Fig.(\ref{fig1}) together with $M=0$ for $v<0$. We observe that,
for any initial mass, $M(v)$ approaches the critical mass $M_{\rm cr}$ for $v\rightarrow \infty$.
This behavior is the most important manifestation of the quantum gravity effects: according to 
Eq.(\ref{nove}), the temperature $T_{\rm BH}(M)$ goes to zero when $M$ approaches $M_{\rm cr}$ from
above. Hence the luminosity vanishes, too, the evaporation process stops, and $M(v)\approx M_{\rm cr}$ remains
approximately constant at very late times, $v \gg M_{\rm cr}^{-1}$. 
%%%%%%%%%%%%%%%%%%%%%%%%%%%%%% FIG 1
In Fig.(\ref{fig1})  we also plot the advanced time dependence of the
temperature $T_{\rm BH}(v)\equiv T_{\rm BH}(M(v))$ and the luminosity $L(v)\equiv L(M(v))$, respectively.
They are obtained by inserting the numerical solution of Eq.(\ref{duetre}) into (\ref{nove})

%%%%%%%%%%%%%%%%%%%%%%%%%%%%%%%%%%%%%%%%%%%%%%%%%%%%%%%%%%%
\begin{figure}
\begin{center}
\includegraphics[width=5cm]{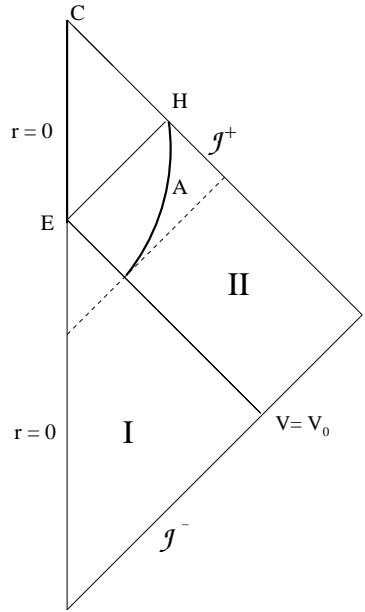}
\caption{\label{figc}
The conformal diagram of the evaporating quantum black hole:   
region I is  a flat spacetime, and 
region II is the evaporating BH spacetime, 
$EH$ is the event horizon, $CH$ is the inner (Cauchy) horizon,
and $A$ is the apparent horizon. }
\end{center}
\end{figure}
%%%%%%%%%%%%%%%%%%%%%%%%%%%%%%%%%%%%%%%%%%%%%%%%%%%%%%%%%%%%%%%
The global structure of the spacetime is depicted in the conformal diagram in
Fig.(\ref{figc}). Region I is a flat spacetime, while at $V=V_0$ ($V$ is the
Kruskal advanced time coordinate, defined as $V =-\exp (-\kappa v)$ being $\kappa$ the surface gravity of the outer horizon)  
an imploding null shell is present (strictly speaking it must have a negative tension  
in order to balance the flux of negative energy on its future side.  
Region II is the evaporating black hole spacetime.  The apparent horizon AH is a
timelike hypersurface which ``meets'' the event horizon EH at  future null
infinity in the conformal diagram. The null ray which is tangent to the earliest portion of the apparent horizon 
$A$ would have been the EH if the hole were not radiating.  
The final state of the black hole is an extremal black hole whose inner 
and outer horizons have the same radius $(r=r_{\rm cr})$ and are located  at 
the event horizon $EH$ and the inner (Cauchy) horizon $CH$ in  Fig.(\ref{figc}).   

It is instructive to compare the areas ${\cal A}$ of the various horizons. 
They are defined by intersecting the EH, AH, and TLS with the 
incoming null surfaces $v= const$. Thus ${\cal A}_{\rm TLS} (v)= 4\pi \rp(v)^2$ and ${\cal A}_{\rm EH}(v)=4\pi\re(v)^2$.
From Eq.(\ref{sette})
we obtain for ${\cal A}_{\rm TLS} \equiv {\cal A}_{\rm AH}$
\be\label{4.12}
{\cal A}_{\rm TLS} = 4\pi G_0^2 M^2 \Big [1+\sqrt{1-(M_{\rm cr}/M)^2}\Big ]^2
\ee 
%and the approximate result (\ref{4.9a}) for the  event horizon implies 
and for the  event horizon implies 
\be\label{4.13}
{\cal A}_{\rm EH}={\cal A}_{\rm TLS} \Big [1-\frac{4\sigma}{(4\pi)^3\widetilde{\omega}}
\frac{\Omega(1-\Omega)}{1+\sqrt{1-\Omega}} \Big ]
\ee
where a term of second order in $\sigma/(4\pi)^3\widetilde{\omega}$ has been neglected. The difference 
$\delta {\cal A}\equiv {\cal A}_{\rm TLS}-{\cal A}_{\rm EH}$ is
given by
\be\label{4.14}
\delta {\cal A}=\frac{\sigma}{4\pi^2} \; \ell_{\rm Pl}^2 \; (1-\Omega) \big [1+\sqrt{1-{\Omega}} \big ]
\ee
During the early stages of the evaporation process, 
$\delta {\cal A}\approx \sigma l_{\rm Pl}^2 /(2\pi^2) = 128 \pi B \ell_{\rm Pl}^2$ 
which coincides with the known result for the Hawking regime, 
while $\delta {\cal A}$ vanishes proportional to $(M^2-M^2_{\rm cr})\rightarrow 0$
for $v \rightarrow \infty$.
It had been emphasized by York 
\cite{1983PhRvD..28.2929Y} that in the Hawking regime he considered, $\delta {\cal A}$ is a
universal ({\it i.e.} $M$ independent) quantity which depends only on $\sigma$, thus counting the degrees of freedom of the 
field quanta which can be evaporated off. Looking at Eq.(\ref{4.14}) we see that this universality does not 
persist beyond the semiclassical approximation.

In conclusion the renormalization group improvement of  black hole spacetimes
according to Quantum Einstein Gravity  leads to concrete predictions on the final state of 
the evaporation process.
Unlike previous studies based on {\it ad hoc} modifications
of the equation of state of matter at very high (Planckian) densities,
or models based on loop quantum gravity,
the mass of the remnant can be calculated explicitly: 
$M_{\rm cr}=\sqrt{\widetilde{\omega}} \ell_{\rm Pl}$. 
Its precise value is  determined by the value of $\widetilde{\omega}$
which is a measurable quantity in principle. 
No naked singularity forms, so that the remnant is a mini-black hole of
Planckian size (See also \cite{2008IJMPD..17..627W} for an approach based on
special resummations of higher order graviton loops, and
\cite{2009PhRvD..79j4009S} for an ``emergent'' spacetime approach).

It is intriguing to note that remnants of this kind of TeV mini-black holes can
have observable signatures at LHC \cite{2006hep.ph....6193B,2005JHEP...06..079R,2008PhRvL.100m1301L}.

\section{Quantum Gravity at astrophysical distances}
The realistic RG trajectory described in Fig.1 terminates before the line
$\lambda=1/2$ as at this point the $\beta$-functions become singular. It is
interesting to see this phenomenon in detail, using for instance the
proper-time formulation of the flow equation \cite{2005JHEP...02..035B},  
%It is convenient to rewrite them in terms of
%the dimensionless Newton constant $g(k)\equiv k^{d-2}G_k\equiv
%k^{d-2}Z_{Nk}^{-1}\bar{G}$ and the dimensionless cosmological constant
%$\lambda(k)\equiv k^{-2}\bar{\lambda}_k$. This leads to the following
%system of equations:
\begin{subequations}
\ba\label{bb}
&&\partial_t g = \beta_g(g,\lambda) \equiv [d-2+\eta_N]g\\[2mm]
&&\partial_t \lambda = \beta_\lambda (g,\lambda)
\ea
\end{subequations}
being $g(k)\equiv k^{d-2}G_k\equiv
k^{d-2}Z_{Nk}^{-1}\bar{G}$ and the dimensionless cosmological constant
$\lambda(k)\equiv k^{-2}\bar{\lambda}_k$. 
The anomalous dimension $\eta_N \equiv -\partial_t {\rm ln} Z_{Nk}$ is given by
\be\label{15}
\eta_N=8(4\pi)^{1-\frac{d}{2}}\Big [ \frac{d(7-5d)}{24}\; (1-2\lambda)^{\frac{d}{2}-m-2}-
\frac{d+6}{6}\Big ]g \;
\frac{\Gamma(m+2-\frac{d}{2})}{\Gamma(m+1)}
\ee
and the beta-function of $\lambda$ reads
\be\label{16}
\beta_\lambda= -(2-\eta_N)\lambda 
+4(4\pi)^{1-\frac{d}{2}}\Big [\frac{d(d+1)}{4}\;(1-2\lambda)^{\frac{d}{2}-m-1}-d\Big ]g
\frac{\Gamma(m+1-\frac{d}{2})}{\Gamma(m+1)}
\ee
where $m>1$ is an integer linked to proper-time regulator
\cite{2005JHEP...02..035B}.

The presence of an IR pole is signaling that the Einstein-Hilbert truncation is
no longer a consistent approximation to the full flow equation, and most
probably a new set of IR-relevant operators is emerging ad $k\rightarrow 0$.
The pole is in fact present in any type  of cutoff in the Einstein-Hilbert
truncation and it is due to the presence of negative eigevalues in the spectrum
of $\Gamma^{(2)}_{k}$. 
As discussed in \cite{2004JCAP...12..001R} the dynamical origin of these strong IR
effect is due to an ``instability driven renormalization'', a phenomenon well known from many other
physical systems \cite{1999PhLB..445..351A,2000PhRvD..62l5021L,2004NuPhB.693...36B}. 

In order to illustrate this point let us look at a scalar model in a simple
truncation:
\begin{align}
\Gamma_{k} [\phi] = \int \!\! \text{d}^{4} x~
\Big \{
\tfrac{1}{2} \, \partial_{\mu} \phi \, \partial^{\mu} \phi
+ \tfrac{1}{2} \, m^{2} (k) \, \phi^{2}
+ \tfrac{1}{12} \, \lambda (k) \, \phi^{4}
\Big \}.
\label{12-2}
\end{align}
Here $\phi$ denotes a real, 
$\mathcal{Z}_{2}$--symmetric scalar field, and the
truncation ansatz \eqref{12-2} retains only a running mass and
$\phi^{4}$--coupling. In a momentum representation we have
\begin{align}
\Gamma^{(2)}_{k} {=} p^{2} + m^{2} (k) + \lambda (k) \, \phi^{2}.
\label{12-3}
\end{align}
Always assuming that $\lambda >0$, we see that $\Gamma^{(2)}_{k}$ is
positive if $m^{2} >0$; but when $m^{2} <0$ it can become negative for
$\phi^{2}$ small enough. Of course, the negative eigenvalue for $\phi
=0$, for example, indicates that the fluctuations want to grow, to
``condense'', and thus to shift the field from the ``false vacuum'' to
the true one. This gives rise to the instability induced renormalizations. 
In fact, the standard $\boldsymbol{\beta}$--functions for $m^{2}$ and $\lambda$
can be found by inserting \eqref{12-3} into the flow equation, taking two and
four derivatives with respect to $\phi$, respectively, and then setting
$\phi=0$ in order to project out $\partial_{t} m^{2}$ and
$\partial_{t} \lambda$. As a result, the
$\boldsymbol{\beta}$--functions are given by $p$--integrals over
(powers of) the propagator
\begin{align}
\Bigl[ p^{2} + m^{2} (k) + k^{2} \Bigr]^{-1}.
\label{12-4}
\end{align}
In the symmetric phase ($m^{2} >0$) this (euclidean!) propagator has
no pole, and the resulting $\boldsymbol{\beta}$--functions are
relatively small. In the broken phase ($m^{2} <0$), however, there is
a pole at $p^{2} = - m (k)^{2} - k^{2}$ provided $k^{2}$ is small
enough: $k^{2} < | m (k)^{2}|$. For $k^{2} \searrow | m (k)^{2}|$ the
$\boldsymbol{\beta}$--functions become large and there are strong
instability induced renormalizations.

In a reliable truncation, a physically realistic RG trajectory in the
spontaneously broken regime will not hit the singularity at $k^{2} = |
m (k)^{2}|$, but rather make $m (k)$ run in precisely such a way that
$| m (k)^{2}|$ is always smaller than $k^{2}$. This requires that
\begin{align}
- m (k)^{2} \propto k^{2}.
\label{12-5}
\end{align}
In order to ``cure" the singularity, a
mass renormalization is necessary in
order to evolve a double-well shaped symmetry breaking classical
potential into an effective potential which is convex and has a flat
bottom.

Unfortunately the two--parameter truncation \eqref{12-2} is too
rudimentary for a reliable description of the broken phase. Its RG
trajectories actually do run into the singularity. They terminate at a
finite scale $k_{\text{term}}$ with $k_{\text{term}}^{2} = | m
(k_{\text{term}})^{2}|$ at which the $\boldsymbol{\beta}$--functions
diverge. Instead, if one allows for an arbitrary running potential
$U_{k} (\phi)$, containing infinitely many couplings, all trajectories
can be continued to $k=0$, and for $k \searrow 0$ one finds indeed the
quadratic mass renormalization \eqref{12-5} \cite{2004NuPhB.693...36B}.

Let us return to gravity now where $\phi$ corresponds to the metric.
In the Einstein--Hilbert truncation it suffices to insert the metric
corresponding to a sphere $\mathrm{S}^{4} (r)$ of arbitrary radius
$r$ into the flow equation
in order to disentangle the
contributions from the two invariants $\int \! \text{d}^{4} x
\sqrt{g\,} \propto r^{4}$ and $\int \! \text{d}^{4} x \sqrt{g\,}
\, R \propto r^{2}$. Thus we may think of the Einstein--Hilbert flow
as being a manifestation of the dynamics of graviton fluctuations on
$\mathrm{S}^{4} (r)$. This family of backgrounds, labeled by $r$, is
``off--shell'' in the sense that $r$ is completely arbitrary and not
fixed by Einstein's equation in terms of $\Lambda$.

It is convenient to decompose the fluctuation $h_{\mu \nu}$ on the
sphere into irreducible (TT, TL, $\cdots$) components \cite{2002PhRvD..65b5013L}
and to expand the irreducible pieces in terms of the corresponding
spherical harmonics. For $h_{\mu \nu}$ in the transverse--traceless
(TT) sector,  the operator $\Gamma^{(2)}_{k} + R_{k}$ equals, up
to a positive constant,
\begin{align}
- D^{2} +  8 \, r^{-2} + k^{2} - 2 \, \Lambda (k)
\label{12-6}
\end{align}
with $D^{2} \equiv g^{\mu \nu} \, D_{\mu} D_{\nu}$ the covariant
Laplacian acting on TT tensors. The spectrum of $- D^{2}$, denoted $\{
p^{2} \}$, is discrete and positive. Obviously \eqref{12-6} is a
positive operator if the cosmological constant is negative. In this
case there are only stable, bounded oscillations, leading to a mild
fluctuation induced renormalization. 
%This is precisely what we observe
%in the IR of the Type Ia trajectories: there is virtually no
%non--canonical parameter running below $k = m_{\text{Pl}}$. 
The situation is very different for $\Lambda > 0$ where, for $k^{2}$
sufficiently small, \eqref{12-6} has negative eigenvalues, i.\,e.\
unstable eigenmodes. In fact, expanding the RHS of the flow equation
to orders $r^{2}$ and $r^{4}$ the resulting
$\boldsymbol{\beta}$--functions are given by traces (spectral sums)
containing the propagator 
\begin{align}
\left[ p^{2} + k^{2} - 2 \, \Lambda (k) \right]^{-1}.
\label{19-1}
\end{align}
The crucial point is that the propagator \eqref{19-1} can have a pole
when $\Lambda (k)$ is too large and positive. It occurs for $\Lambda
(k) \geq k^{2}/2$, or equivalently $\lambda (k) \geq 1/2$, at $p^{2} =
2 \, \Lambda (k) - k^{2}$. Upon performing the $p^{2}$--sum this pole
is seen to be responsible for the terms $\propto 1 / \left( 1 - 2 \,
  \lambda \right)$ and $\ln (1 - 2 \, \lambda)$ in the
$\boldsymbol{\beta}$--functions which become singular at $\lambda =
1/2$. The allowed part of the $g$-$\lambda$--plane ($\lambda < 1/2$)
corresponds to the situation $k^{2} > 2 \,
\Lambda (k)$ where the singularity is avoided thanks to the large
regulator mass. When $k^{2}$ approaches $2 \, \Lambda (k)$ from above
the $\boldsymbol{\beta}$--functions become large and strong
renormalizations set in, driven by the modes which would go
unstable at $k^{2} = 2 \,\Lambda$.

In this respect the situation is completely analogous to the scalar
theory discussed above: Its symmetric phase ($m^{2} >0$) corresponds
to gravity with $\Lambda <0$; in this case all fluctuation modes are
stable and only small renormalization effects occur. Conversely, in
the broken phase ($m^{2} <0$) and in gravity with $\Lambda >0$, there
are modes which are unstable in absence of the IR regulator. They lead
to strong IR renormalization effects for $k^{2} \searrow | m (k)^{2}|$
and $k^{2} \searrow 2 \, \Lambda (k)$, respectively. The gravitational
Type Ia (Type IIIa) trajectories are analogous to those of the
symmetric (broken) phase of the scalar model.

In view of the scalar analogy it is a plausible and very intriguing
speculation that, for $k \to 0$, an improved gravitational truncation
has a similar impact on the RG flow as it has in the scalar case.
There the most important renormalization effect is the running of the
mass: $- m (k)^{2} \propto k^{2}$. If gravity avoids the singularity
in an analogous fashion the cosmological constant would run
proportional to $k^{2}$,
\begin{align}
\Lambda (k) = \lambda^{\text{IR}}_{*} \, k^{2}
\label{19-2}
\end{align}
with a constant $\lambda^{\text{IR}}_{*} < 1/2$. In dimensionless
units \eqref{19-2} reads $\lambda (k) = \lambda^{\text{IR}}_{*}$,
i.\,e.\ $\lambda^{\text{IR}}_{*}$ {\it is a infrared fixed point} of the
$\lambda$--evolution. If the behavior \eqref{19-2} is actually
realized, the renormalized cosmological constant observed at very
large distances, $\Lambda (k \to 0)$, vanishes regardless of its bare
value. 
%Clearly this would have an important impact on the cosmological
%constant problem \cite{1989RvMP...61....1W,1987PhLB..188...38R}.

%$k\rightarrow 0$ both $g$ and $\lambda$ run into an IR attractive
%non-Gaussian fixed
%point, {\it i.e.} that for a wide range of initial conditions
%\be\label{6}
%\lim_{k\rightarrow 0} g(k) = \gir, \;\;\;\; \lim_{k\rightarrow 0}\lambda(k) =
%\lir
%\ee
%where $\gir$ and $\lir$ are strictly positive. 

The above discussion has thus lead to the conjecture that the IR behavior of the 
Newton constant and the cosmological constant is regulated by an IR attractive
fixed point. Several investigations 
\cite{2002PhLB..527....9B,2004IJMPD..13..107B,2004JCAP...01..001B,2006CQGra..23.3103B,2004ForPh..52..650R}
have shown that in this framework  a solution of the ``cosmic coincidence
problem'' arises naturally without the introduction of a {\rm quintessence} field. 
In particular in the fixed point regime the
vacuum energy density $\rho_\Lambda\equiv\Lambda/8\pi G$
is automatically adjusted so as to equal the matter energy density,
{\it i.e.} $\OL=\OM =1/2$, and that the deceleration parameter approaches $q =
-1/4$. Moreover, an analysis of the high-redshift SNe Ia data leads to the conclusion that
this {\em infrared fixed point cosmology} is in good  agreement with
the observations \cite{2004JCAP...01..001B}.

More recent works have instead considered the possibility that the ``basin of
attraction'' of the IR fixed point can act already at galactic scale, thus
providing an explanation for the galaxy rotation curve without dark matter 
\cite{2004JCAP...12..001R,2004PhRvD..69j4022R,2004PhRvD..70l4028R,2007CQGra..24.6255E},
but a detailed analysis based on available experimental data is still missing.

In conclusion, although the existence of an {\it infrared fixed point} can only
be conjectured on the basis of the above argument, the RG cosmologies derived
from it are promising candidates to explain the Dark Energy and Dark Matter
issue.

\section{Conclusions}
%%%%%%%%%%%%%%%%%%%%%%%%%%%%%%%%%%%%%%%%%%%%%%%%%%%%%%%%%%%%%%%%%%%%%%%%%%%%%%%%%%%

In these notes some important astrophysical consequences of the Asymptotic
Safety Scenario have been reviewed. 

In particular
it was advocated the point of view that the scale dependence of
the gravitational parameters has an impact on the physics of the Universe we
live in and, in particular, it has been possible to identify known
features of the Universe which could possibly  be due to this scale dependence.
Three possible candidates for such features are proposed: the entropy
carried by the radiation which fills the Universe today, 
a period of automatic, $\Lambda$-driven inflation 
that requires no ad hoc inflaton, and the primordial density perturbations. 

Moreover, the impact of the leading quantum gravity effects on the dynamics of
the Hawking evaporation process of a black hole have also been investigated. 
Its spacetime structure is described by a renormalization group improved Vaidya
metric. Its event horizon, apparent horizon, and timelike limit surface 
have been obtained taking the scale dependence of Newton's constant into
account. The emergence of a quantum
ergosphere is discussed. 
The final state of the evaporation process is a cold, Planck size remnant.

It would be interesting
to investigate the possible astrophysical implications of a population  of
stable Planck size mini-black holes produced in the Early Universe or by the interaction of
cosmic rays with the interstellar medium. I hope to address this issue in
a subsequent publication.

\section{Acknowledgments}
It is a pleasure  to thank Daniel Litim and all the organizers of the Brighton
Workshop on Continuum and Lattice Approaches to Quantum Gravity for their
cordial hospitality and for creating a stimulating scientific atmosphere.
I would also like to thank Martin Reuter for useful comments. 

\bibliography{t2}

%\end{thebibliography}
\end{document}